\documentclass[preprint,12pt,authoryear]{elsarticle}

\usepackage{onimage}
\usepackage{geometry}
\usepackage{amssymb,amsmath,amsfonts,empheq}
\usepackage{lineno}
\usepackage{tikz,calc}
\usepackage{pgfplots}
\usetikzlibrary{decorations.pathmorphing,arrows}
\usepackage{tabularx}
\usepackage{nohyperref} 
\usepackage{url} 
\usepackage{algorithm,algpseudocode}

\tikzset{snake it/.style={decorate, decoration=snake}}

\newcommand{\Ical}{\mathcal{I}}
\newcommand{\Icalp}{\Ical^{(p)}}
\newcommand{\Icalf}{\Ical^{(f)}}
\newcommand{\Vcal}{\mathcal{V}}
\newcommand{\Vcalp}{\Vcal^{(p)}}
\newcommand{\Vcalf}{\Vcal^{(f)}}
\newcommand{\Scal}{\mathcal{S}}
\newcommand{\Scalp}{\Scal^{(p)}}
\newcommand{\Scalf}{\Scal^{(f)}}

\newcommand{\avgrhokg}{\langle \rho_k \rangle^{(g)}}
\newcommand{\avgrhog}{\langle \rho \rangle^{(g)}}
\newcommand{\avgrhos}{\langle \rho \rangle^{(s)}}
\newcommand{\avgukig}{\langle u_{k,i} \rangle^{(g)}}
\newcommand{\avguig}{\langle u_i \rangle^{(g)}}

\newcommand{\avgeg}{\langle e \rangle^{(g)}}
\newcommand{\avghg}{\langle h \rangle^{(g)}}
\newcommand{\avghs}{\langle h \rangle^{(s)}}

\newcommand{\avgT}{\langle T \rangle}

\newcommand{\vn}{\mathbf{n}}

\newcommand{\vx}{\mathbf{x}}

\newcommand{\Bg}{B^\prime_g}
\newcommand{\Bc}{B^\prime_c}
\newcommand{\Bf}{B^\prime_{fl}}

\newcommand{\ddt}{\frac{d}{d t}}

\newcommand{\revision}[1]{\textcolor{black}{#1}}
\newcommand{\revisiontwo}[1]{\textcolor{black}{#1}}

\pgfplotsset{compat=1.18} 
\begin{document}
\begin{frontmatter}



\title{An Extended $B^{\prime}$ Formulation for Ablating-Surface Boundary Conditions}

\author[label1,label3]{Alberto Padovan\corref{cor1}}
\ead{padovan3@illinois.edu}
\cortext[cor1]{Corresponding author}
\author[label1,label3]{Blaine Vollmer}
\author[label1,label3]{Francesco Panerai}
\author[label1,label3]{Marco Panesi}
\author[label2,label3]{Kelly A. Stephani}
\author[label1,label3]{Daniel J. Bodony}

\affiliation[label1]{organization={Department of Aerospace Engineering, University of Illinois at Urbana-Champaign},
            addressline={104 S. Wright St.}, 
            city={Urbana},
            postcode={61801}, 
            state={IL},
            country={USA}}
            
\affiliation[label3]{organization={Center for Hypersonics \& Entry Systems Studies, University of Illinois at Urbana-Champaign},
            addressline={105 S. Goodwin Ave.}, 
            city={Urbana},
            postcode={61801}, 
            state={IL},
            country={USA}}
            
\affiliation[label2]{organization={Department of Mechanical Science and Engineering, University of Illinois at Urbana-Champaign},
            addressline={1206 W. Green St.}, 
            city={Urbana},
            postcode={61801}, 
            state={IL},
            country={USA}}

\begin{abstract}

The $B^\prime$ formulation can be understood as a mass and energy conservation formalism at a reacting singular surface. 
In hypersonics applications, it is typically used to compute the chemical equilibrium properties of gaseous mixtures at ablating surfaces, and to estimate the recession velocity of the interface. 
In the first half of the paper, we derive the $B^\prime$ formulation to emphasize first principles.
\revision{In particular, while we eventually specialize to the commonly considered case of chemical equilibrium boundary layers that satisfy the heat and mass transfer analogy, we first derive a general interface jump condition that lets us highlight all the underlying assumptions of the well-known $B^\prime$ equations.}
This procedure helps elucidate the nature of the $B^\prime$ formalism and it also allows us to straightforwardly extend the original formulation.
Specifically, when applied at the interface between a porous material and a boundary layer (as in thermal protection systems applications), the original formulation assumes unidirectional advective transport of gaseous species from the porous material to the boundary layer (i.e., blowing). 
However, under conditions that may appear in hypersonic flight or in ground-based wind tunnels, boundary layer gases can enter the porous material due to a favorable pressure gradient. 
We show that this scenario can be easily handled via a straightforward modification to the $B^\prime$ formalism, and we demonstrate via examples that accounting for gas entering the material can impact the predicted recession velocity of ablating surfaces. 
\revision{In order to facilitate the implementation of the extended $B^\prime$ formulation in existing material response codes, we present a short algorithm in section \ref{sec:extension_bprime} and we also refer readers to a GitHub repository where the scripts used to generate the modified $B^\prime$ tables are publicly available.}
\end{abstract}


\end{frontmatter}


\section{Introduction}

Understanding the fluid-structure interaction between a high-speed boundary layer and a reacting porous material is important for various applications, including the design of thermal protection systems (TPS) for atmospheric reentry. 
High-fidelity simulations that aim to study the coupled physics between the fluid and the solid, necessarily require access to computational fluid dynamics (CFD) codes that simulate the physics of the boundary layer, and to material response codes that simulate the dynamic response of the material. 
However, if we are \revision{primarily} interested in studying the response of the material, \revision{or if we seek a low-resolution estimate of the fluid-material interaction}, fully resolving the boundary layer dynamics is a computational burden. 
\revision{In order to circumvent the need to perform a fully resolved CFD calculation, researchers have developed first-principles formulations that model the mass, momentum and energy transfer at the interface between a reacting solid and a boundary layer.
The $B^\prime$ formalism discussed in this paper is one such formulation, and it allows to (i) run the material response code independently of a fluid solver when we are uninterested in the fluid mechanics, or (ii) provide a low-resolution interface boundary condition when we seek a low-resolution estimate of the coupled system.}

The $B^\prime$ formalism can be considered as a mass and energy flux-balance condition, arising from a control volume analysis at the interface between two different media. 
In TPS and ablation applications \citep{moyer1970,moyer1970_2}, where the interface separates a high-speed boundary layer from a chemically-reacting porous material, this formalism is needed to estimate the surface recession velocity under the assumption of chemical equilibrium at the interface.
The convenience of the formulation lies in its computational simplicity, and in the fact that, under \revision{several assumptions discussed in sections \ref{sec:cons_mass} and \ref{sec:cons_energy}}, the solution of the $B^\prime$ equation can be tabulated as a function of surface temperature, surface pressure and normalized gas mass flux (hence the common name \emph{$B^\prime$ tables}). 

Although the original $B^\prime$ formulation is known and implemented in ablation codes (e.g., PATO \citep{lachaud2014} and KATS \citep{weng2014}), to the best of the authors' knowledge, a derivation from first principles is not readily available in the literature. 
Specifically, the original formulation is typically presented starting from an infinitesimally thin control volume containing the interface \citep{moyer1968,moyer1970_2,anderson1970,demuelenaere2012,lachaud2014,bellas2018}.
In sections \ref{sec:jump_cond}, \ref{sec:cons_mass} and \ref{sec:cons_energy} we offer \revisiontwo{an alternative derivation} of the $B^\prime$ mass and energy balance equations starting from a \emph{jump condition} that is derived using the divergence and the generalized transport theorems \citep{keller1954}, without explicitly requiring an infinitesimally thin control volume.
\revision{While in sections \ref{sec:cons_mass} and \ref{sec:cons_energy} we specialize to boundary layers with unity Lewis numbers (as commonly done in the literature), the jump condition presented in section \ref{sec:jump_cond} is general enough that it can be applied to any boundary layer model.
This way, we elucidate the nature of the~$B^\prime$ formalism and identify the underlying assumptions that are built into it. 
}


In section \ref{sec:extension_bprime} we use the derivation presented in the first half of the manuscript to extend the $B^\prime$ formulation to include bidirectional mass flux across the interface.
In its original form, the $B^\prime$ formulation assumes unidirectional advective transport of gaseous mass from the porous material to the boundary layer (i.e., blowing). 
(This corresponds to $\Bg > 0$ in the notation of section \ref{sec:cons_mass}.)
This is because the formulation is often used to simulate the response of pyrolyzing porous materials \citep{moyer1970,lachaud2014,chiodi2022} that exhibit internal pressures that are often higher than the pressure inside the boundary layer (thereby leading to blowing). 
However, there can be cases where, even in the presence of pyrolysis, the pressure differential is such that there is a net inflow of gases into the porous material ($\Bg < 0$). 
\revisiontwo{In computational codes that treat the porous material's gases as a time-varying (equilibrium/non-equilibrium) mixture, the inflow of boundary layer gases into the material can be easily accounted for via a species dirichlet boundary condition at the surface \citep{lachaud2015}.
Conversely, when the gases composition is taken to be constant (i.e., when there is no species tracking), existing codes (e.g., PATO \citep{lachaud2014}, CHyPS \citep{chiodi2022}) typically choose to neglect the effect of the inflow of gases on the surface thermodynamics by setting $\Bg = 0$.
We shall see, however, that enforcing $\Bg = 0$ can have a non-negligible effect on the surface thermodynamics and on the surface recession velocity.}
Section \ref{sec:extension_bprime} presents an extension to the $B^\prime$ formalism that allows for $\Bg < 0$ \revisiontwo{even when the gases in the porous material are treated as a constant mixture}.
First and foremost, this extension has the same computational cost as the original formulation and, just like the latter, it allows for the $B^\prime$ equations to be tabulated a priori. 
Second, it is constructed such that the normalized recession rate ($\Bc$ in the notation of section \ref{sec:cons_mass}) is a continuous function of the blowing/\revision{aspiration} rate~$\Bf$. 
Finally, we identify blowing/aspiration regimes where the recession rate is either independent of the blowing/aspiration rate, or a linear function of the latter. 
This analysis shows that if the mass flux of gases into the porous material is sufficiently high, its impact on the surface recession velocity is non-negligible. 
This is demonstrated via examples in section \ref{sec:results}, where we show that the modified formulation predicts recession velocities that are always equal to or greater than the recession velocities predicted by the classical $B^\prime$ formulation. 

\revision{The steps required to implement this formulation in existing material response codes are compactly outlined in Algorithm \ref{alg:bprime} in section \ref{sec:extension_bprime}. 
Moreover, the interested reader may generate the modified $B^\prime$ tables using the scripts that are publicly available in the repository \url{https://github.com/albertopadovan/Modified_Bprime}. 
The $B^\prime$ tables generated by these scripts should be compatible with the material response code CHyPS \citep{chiodi2022} without modification, and with PATO \citep{lachaud2014} with little to no modification. 
}

\section{Jump Condition of a Conserved Quantity}
\label{sec:jump_cond}

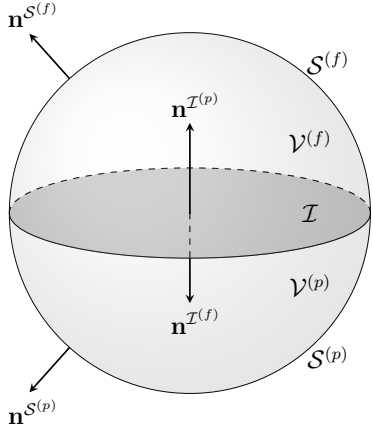
\begin{figure}
    \centering
    \scalebox{0.8}{
    \begin{tikzpicture}
        \filldraw[fill=gray,opacity=0.4] (-3,0) arc (180:360:3cm and 0.75cm);
        \filldraw[dashed,fill=gray,opacity=0.4] (-3,0) arc (180:0:3cm and 0.75cm);
        \draw (-3,0) arc (180:360:3cm and 0.75cm);
        \draw[dashed] (-3,0) arc (180:0:3cm and 0.75cm);
        \draw (0,0) circle (3cm);
        \shade[ball color=blue!10!white,opacity=0.15] (0,0) circle (3cm);
        \filldraw (0,0) circle[radius=0.2pt];
        \draw[-stealth,thick] (0,0) -- (0,1.5);
        \draw[dashed] (0,0) -- (0,-0.75);
        \draw[-stealth,thick] (0,-0.75) -- (0,-1.5);
        \node at (0.1,1.8){$\mathbf{n}^{\Icalp}$};
        \node at (0.1,-1.8){$\mathbf{n}^{\Icalf}$};
        \node at (2,0) {$\Ical$};
        \node at (2,1.2) {$\Vcalf$};
        \node at (2,-1.2) {$\Vcalp$};
        \node at (2.3,2.5) {$\Scalf$};
        \node at (2.3,-2.4) {$\Scalp$};
        \draw[-stealth,thick] (-2,2.23606797749979) -- (-2.66666667,2.98142397);
        \node at (-2.6,3.3) {$\mathbf{n}^{\Scalf}$};
        \draw[-stealth,thick] (-2,-2.23606797749979) -- (-2.66666667,-2.98142397);
        \node at (-2.6,-3.3) {$\mathbf{n}^{\Scalp}$};
    \end{tikzpicture}}
    \caption{Schematic of a general control volume $\Vcal = \Vcalf\cup \Vcalp$ containing the interface $\Ical$ between a porous material and a fluid. Superscripts $(f)$ and $(p)$ denote the fluid and porous material's sides, respectively. A description of the variables is given at the beginning of section \ref{sec:jump_cond}.} 
    \label{fig:ctrl_vol}
\end{figure}

In this section we follow the approach of \cite{keller1954} to derive the jump condition of a conserved quantity $\varphi$ across a singular surface where $\varphi$ is discontinuous. 
Throughout, we use the general control volume $\Vcal = \Vcal^{(p)}\cup \Vcal^{(f)} \subseteq \mathbb{R}^3$ depicted in figure~\ref{fig:ctrl_vol}, where the superscripts $(p)$ and $(f)$ denote the two sides of the control volume that are separated by the singular interface $\Ical$. 
In applications of interest, the $(p)$-side will contain a volume of porous material, while the $(f)$-side will contain a volume of fluid. 
The external surfaces of the control volume are denoted by $\Scal$, and~$\vn\in\mathbb{R}^3$ denotes the unit-norm outward-pointing vector normal to the surface.

In multi-physics problems, the spatio-temporal dynamics of conserved quantities are typically governed by partial differential equations defined on either side of the interface $\Ical$. 
For a general quantity $\varphi(\vx,t)$, the conservation equations in differential conservative form (and Einstein notation) may read
\begin{empheq}[left = \empheqlbrace]{align}
    \frac{\partial \varphi^{(f)}}{\partial t} + \frac{\partial}{\partial x_i}\left(\varphi^{(f)} u_i^{(f)}\right) &= \frac{\partial}{\partial x_i}\xi^{(f)}_i +\psi^{(f)},\quad \vx\in\Vcalf,\label{eq:cons_diff_phif}\\
    \frac{\partial \varphi^{(p)}}{\partial t} + \frac{\partial}{\partial x_i}\left(\varphi^{(p)} u_i^{(p)}\right) &= \frac{\partial}{\partial x_i}\xi^{(p)}_i +\psi^{(p)},\quad \vx\in\Vcalp.
    \label{eq:cons_diff_phip}
\end{empheq}
Here, $u_i$ denotes the $i$th component of the transport velocity vector, $\psi$ denotes a volumetric source term and $\xi_i$ denotes the $i$th component of additional terms (e.g., the viscous stress tensor in the momentum equation, or viscous dissipation in the energy equation).
Equations \eqref{eq:cons_diff_phif} and \eqref{eq:cons_diff_phip} are well-posed on $\Vcalf$ and $\Vcalp$, respectively, where~$\varphi$ is differentiable with respect to~$\vx$, but they do not hold \emph{on} the interface, where~$\varphi$ typically exhibits a discontinuity.
Understanding this discontinuity, and deriving the corresponding jump condition, is at the heart of imposing the correct boundary conditions in computational codes that run multi-physics simulations.

In order to derive the jump condition, we turn to the integral form of the conservation equations. 
In particular, the conservation equation over $\Vcalf$ is given by
\begin{equation}
\label{eq:cons_varphif}
    \begin{aligned}
        &\ddt \int_{\Vcalf}\varphi^{(f)}d\Vcalf + \int_{\Scalf}\left[\varphi^{(f)}\left(u^{(f)}_{i} - v_i^{\Scalf}\right)-\xi_i^{(f)}\right]n_i^{\Scalf}d\Scalf \\
        &+ \int_{\Ical}\left[\varphi^{(f)}\left(u^{(f)}_{i} - v_i^{\Ical}\right)-\xi_i^{(f)}\right] n_i^{\Icalf}d\Ical = \int_{\Vcalf} \psi^{(f)}d\Vcalf,
    \end{aligned}
\end{equation}
where $v_i$ is the $i$th component of the surface velocity vector, and $n_i$ is the $i$th component of the outward-pointing normal vector.
Using the generalized transport theorem on the time-rate-of-change term in \eqref{eq:cons_varphif}, and making use of the divergence theorem, it can be checked that equations \eqref{eq:cons_varphif} and \eqref{eq:cons_diff_phif} are indeed equivalent.
The conservation equation over $\Vcalp$ is analogous to \eqref{eq:cons_varphif}, with superscripts~$(p)$.

We proceed by considering the integral form of the conservation equation for~$\varphi$ over the whole control volume $\Vcal$,
\begin{equation}
\label{eq:cons_varphi}
    \begin{aligned}
    &\ddt \bigg\{\int_{\Vcalf} \varphi^{(f)} d\Vcalf + \int_{\Vcalp} \varphi^{(p)} d\Vcalp\bigg\} + \int_{\Scalf}\left[\varphi^{(f)}\left(u^{(f)}_{i} - v_i^{\Scalf}\right)-\xi_i^{(f)}\right]n_i^{\Scalf}d\Scalf \\
    &+ \int_{\Scalp}\left[\varphi^{(p)}\left(u^{(p)}_{i} - v_i^{\Scalp}\right)-\xi_i^{(p)}\right]n_i^{\Scalp}d\Scalp = \int_{\Vcalf}\psi^{(f)} d\Vcalf  \\ 
    &+\int_{\Vcalp}\psi^{(p)} d\Vcalp + \int_\Ical \psi^{\Ical} d\Ical.
    \end{aligned}
\end{equation}
In writing equation \eqref{eq:cons_varphi}, we make two assumptions. 
First, we do not allow for any accumulation of quantity $\varphi$ on the interface $\Ical$. 
(This would appear as the time-rate of change of the surface integral of $\varphi$ along $\Ical$.)
Second, we treat $\Ical$ as a reactive interface, which is allowed to create/destroy some amount of $\varphi$ via the surface source term $\psi^\Ical$.
These are modelling assumptions that can, in principle, be relaxed.
\revision{For instance, an example of a more involved interface model can be found in \cite{whitaker1992}, where the author considers a finite-thickness interface that is allowed to accumulate mass.}
Subtracting formula \eqref{eq:cons_varphif} and its analog over $\Vcalp$ from \eqref{eq:cons_varphi}, and imposing point-wise equality, the desired jump condition reads
\begin{equation}
\label{eq:jump_cond_varphi}
    \left[\varphi^{(f)}\left(u_i^{(f)} - v_i\right)-\xi_i^{(f)}\right]n_i - \left[\varphi^{(p)}\left(u_i^{(p)} - v_i\right)n_i -\xi_i^{(p)}\right]n_i = \psi^\Ical,
\end{equation}
where we have used $n_i \coloneqq n_i^{\Icalp} = -n_i^{\Icalf}$, and we have dropped the superscript $\Ical$ on~$v_i$ for notational simplicity. 
In the next sections, we will use \eqref{eq:jump_cond_varphi} to derive the mass and energy jump conditions at an ablating surface. 

\section{$B^\prime$ Formulation from First Principles: Conservation of Mass}
\label{sec:cons_mass}

We use the results from the previous section to derive the well-known $B^\prime$ mass balance equation. 
In doing so, we elucidate the nature of the $B^\prime$ formulation and we identify all its underlying assumptions.

\subsection{Conservation of mass at an ablating surface}
\label{subsec:cons_mass_abl_surf}

Moving forward, we specialize to the case of an ablating surface at the interface between a porous material and a fluid. 
We let the porous material occupy the $\Vcalp$ region of the control volume in figure \ref{fig:ctrl_vol}, while the fluid occupies the $\Vcalf$ side.
If the fluid is a reacting mixture of $N_s$ species, the differential form of the continuity equation for species $k$ is given by 
\begin{equation}
\label{eq:cons_mass_k}
    \frac{\partial \rho_k^{(f)}}{\partial t} + \frac{\partial}{\partial x_i}\left(\rho_k^{(f)} u_{k,i}^{(f)}\right) = \psi_k^{(f)},\quad k\in \{1,2,\ldots,N_s\},
\end{equation}
where $\rho_k$ and $u_{k,i}$ are the density and $i$th component of the velocity associated with species $k$, and $\psi_k$ is a volumetric source term due to the reacting nature of the mixture. 
For future reference, we also define the mixture density $\rho^{(f)}$ and the mixture bulk velocity $u_i^{(f)}$ by \citep{eckert1969}
\begin{equation}
\label{eq:bulk_quantities_f}
    \rho^{(f)} = \sum_{k=1}^{N_s} \rho_k^{(f)},\quad u_i^{(f)} = \frac{1}{\rho^{(f)}}\sum_{k=1}^{N_s} \rho_k^{(f)}u_{k,i}^{(f)}.
\end{equation}


The governing equations for the porous material will be treated in a volume-averaged sense. 
Let the porous material be made of a solid phase and a gaseous mixture with $N_s$ species.
In applications of interest, the porous material is typically made up of several solid phases, but for the current discussion it suffices to consider one. 
Additional solid phases can be considered with minimal change. 
Conservation of mass of species $k$ within the porous material requires that equation \eqref{eq:cons_mass_k} be satisfied (with all superscripts $(f)$ converted to $(p)$), where the source term $\psi_k^{(p)}$ may now account for both homogeneous and heterogeneous reactions. 
The averaging theorem of \cite{whitaker1967} and the modified averaging theorem of \cite{gray1975} allow us to volume-average equation \eqref{eq:cons_mass_k} over a representative elemental volume $V$ to obtain 
\begin{equation}
\label{eq:cons_mass_k_volavg}
    \frac{\partial}{\partial t}\left(\varepsilon_g \avgrhokg\right) + \frac{\partial}{\partial x_i}\left(\varepsilon_g \avgrhokg \avgukig\right) = \langle \psi_k\rangle^{(g)}.
\end{equation}
Here, $\varepsilon_g$ is the volume fraction occupied by the mixture within the representative volume $V$, and $\avgrhokg $ is the \emph{intrinsic} volume average of $\rho_k$ \citep{gray1976}.
In the interest of clarity, we stress that the representative elemental volume $V$ is not related to $\Vcal$ in figure \ref{fig:ctrl_vol}. 
A schematic of $V$ can be found, for instance, in \cite{gray1976}.
It is also important to remark that \eqref{eq:cons_mass_k_volavg} is not exact. 
In fact, the averaging procedure leads to unclosed terms that are typically neglected, either due to physically-justifiable reasons, or to the impossibility of properly closing them (see equation (24) in \cite{gray1976}). 
Once again, for future reference, we let~$\avgrhog$ and $\avguig$ be the volume-averaged mixture density and mixture bulk velocity, defined analogously to \eqref{eq:bulk_quantities_f}.
Finally, the volume-averaged conservation of solid mass reads
\begin{equation}
\label{eq:cons_mass_solid_avg}
    \frac{\partial}{\partial t}\left(\varepsilon_s \avgrhos\right) = \langle \psi_s\rangle^{(s)},
\end{equation}
where $\varepsilon_s = 1-\varepsilon_g$ is the volume fraction occupied by the solid. 
In order to guarantee that, within $\Vcalp$, the sum of mixture mass and solid mass is conserved in the absence of mass fluxes through the boundaries, the source terms are usually taken to satisfy
\begin{equation}
\label{eq:constraint_source_terms}
    \langle\psi_s\rangle^{(s)} + \sum_{k=1}^{N_s}\langle\psi_k\rangle^{(g)} = 0. 
\end{equation}

\subsubsection{Conservation of mass of gaseous species $k$}
\label{subsubsec:cons_mass_species_k}
We now return to our control volume $\Vcal$ in figure \ref{fig:ctrl_vol}. 
Per our previous discussion, conservation of mass of species $k$ in the $\Vcalf$ region of the control volume is governed by \eqref{eq:cons_mass_k}, while conservation of mass of species $k$ in the $\Vcalp$ region is governed in a volume-averaged sense by \eqref{eq:cons_mass_k_volavg}.
The jump condition in \eqref{eq:jump_cond_varphi} can be used directly, and it reads
\begin{equation}
\label{eq:jump_condition_species_k}
    \rho_k^{(f)}\left(u_{k,i}^{(f)}-v_i\right)n_i - \varepsilon_g \avgrhokg \left(\avgukig - v_i\right)n_i = \psi_k^\Ical,
\end{equation}
where $\psi_k^\Ical$ is the rate of production (per unit area) of species $k$ due to reactions at the interface. 
In ablation applications, this production term models the heterogeneous reactions through which the solid phase of the porous material is converted into gaseous mass (thereby causing surface recession). 
This will become clear in the next section \ref{subsubsec:cons_solid_mass}. 

\subsubsection{Conservation of solid mass}
\label{subsubsec:cons_solid_mass}

As in the previous section \ref{subsubsec:cons_mass_species_k}, we can apply the interface balance equation~\eqref{eq:jump_cond_varphi} directly. 
Since there is no solid phase in the $\Vcalf$ region of the control volume $\Vcal$, and~\eqref{eq:cons_mass_solid_avg} governs the volume-averaged continuity of solid mass in the $\Vcalp$ region, equation \eqref{eq:jump_cond_varphi} reduces to 
\begin{equation}
\label{eq:jump_cond_solid_mass}
    \varepsilon_s\avgrhos v_i n_i = \psi_s^\Ical.
\end{equation}
This equation states that the surface velocity $v_i n_i$ of the interface $\Ical$ is proportional to~$\psi_s^\Ical$, where, in ablation applications, $\psi_s^\Ical$ can be understood as the time-rate of change per unit area of solid mass lost to gaseous mass via heterogeneous reactions.
As a sanity check, if solid mass is being lost to gaseous mass (e.g., during ablation), then $\psi_s^{\Ical} < 0$, so $v_i n_i < 0$. 
Since by convention $n_i = n_i^{\Icalp}$, this means that the surface is receding (see figure \ref{fig:ctrl_vol}), as expected. 



\subsection{The $B^\prime$ mass balance}
\label{subsec:bprime_mass_balance}

The $B^\prime$ equation for mass conservation is derived from \eqref{eq:jump_condition_species_k} after a number of assumptions that we will outline shortly.
Before proceeding we remark that the assumptions outlined herein may or may not be physically justified. 
We are merely making them in order to obtain the $B^\prime$ mass balance equation from \eqref{eq:jump_condition_species_k}.

By adding and subtracting $\rho_k^{(f)}u_{i}^{(f)}$ and $\varepsilon_g\avgrhokg\avguig$ to \eqref{eq:jump_condition_species_k}, and using the fact that $\rho_k = z_{k}\rho$, where $z_k$ is the mass fraction of species~$k$, equation \eqref{eq:jump_condition_species_k} can be written as 
\begin{equation}
\label{eq:jump_cond_reshuffled}
    J_{k,i}^{(f)}n_i + z_k^{(f)}\rho^{(f)}\left(u_i^{(f)}-v_i\right)n_i = J_{k,i}^{(g)}n_i + \varepsilon_g z_k^{(g)}\avgrhog\left(\avgukig - v_i\right)n_i + \psi_k^\Ical,
\end{equation}
where $J_{k,i}$ are mass diffusion terms defined as
\begin{equation}
\label{eq:defn_J}
    J_{k,i}^{(f)} = \rho_k^{(f)}\left(u_{k,i}^{(f)}-u_i^{(f)}\right),\quad J_{k,i}^{(g)} = \varepsilon_g\avgrhokg\left(\avgukig - \avguig\right).
\end{equation}
In order to arrive at the well-known $B^\prime$ equation, the following assumptions need to be made. 
First, mass diffusion on the porous material's side of the interface (i.e., $J_{k,i}^{(p)}n_i$) is neglected. 
The mass diffusion term on the fluid's side of the interface is modelled via correlation (or transfer potential) as $J_{k,i}^{(f)} = \rho_e u_{e,i} St_M \left(z_k^{(f)} - z_k^{(e)}\right)$, where \revision{$St_M$ is the mass-transfer Stanton number}, and the subscript/superscript ``e" denotes boundary layer edge quantities \citep{eckert1969}.
\revision{While more detailed mass diffusion models can be considered \citep{kendall1968,lachaud2017}, the transfer potential model considered here is the simplest, and it relies on the assumption that all species share the same mass diffusion coefficient (see also \ref{app:St}).}
Putting this all together, equation \eqref{eq:jump_cond_reshuffled} becomes
\begin{equation}
\label{eq:jump_cond_reshuffled2}
    \begin{aligned}
    &\rho_e u_{e,i} St_M \left(z_k^{(f)} - z_k^{(e)}\right)n_i + z_k^{(f)}\underbrace{\rho^{(f)}\left(u_i^{(f)}-v_i\right)n_i}_{\coloneqq\dot{m}^{(f)}} \\ &= z_k^{(g)}\underbrace{\varepsilon_g \avgrhog\left(\avguig-v_i\right) n_i}_{\coloneqq\dot{m}^{(g)}} + \psi_k^\Ical.
    \end{aligned}
\end{equation}

\revision{In order to obtain the $B^\prime$ equation that is commonly presented in the literature (and implemented in computational codes), we first need to convert \eqref{eq:jump_cond_reshuffled2} to its analog in terms of \emph{elements} rather than species.
Under the assumption of equal diffusion coefficients (so that the definition of $St_M$ remains unchanged), it is straightforward to see that equation \eqref{eq:jump_cond_reshuffled2} can be transformed into
\begin{equation}
    \rho_e u_{e,i} St_M \left(y_k^{(f)} - y_k^{(e)}\right)n_i + y_k^{(f)}\dot{m}^{(f)} = y_k^{(g)}\dot{m}^{(g)} + \chi^{\Ical}_{k},\quad k \in \{1,2,\ldots,N_{es}\},
\end{equation}
where $y_k$ is the mass fraction of element $k$ in the mixture, $N_{es}$ is the number of elements, and $\chi_k^{\Ical}$ is the surface source term analogous to $\psi_k^\Ical$.
At this point we are ready to make the final assumption that ultimately leads to the $B^\prime$ equation. 
Specifically, we write the source term $\chi_k^{\Ical}$ as $\chi_k^{\Ical} = \chi_{k_C}^{\Ical} \delta_{k,k_C}$, where $\delta_{k,k_C}$ is the Kronecker delta and $k_C \in \{1,2,\ldots,N_{es}\}$ is the index pointing to monatomic carbon gas. 
Physically, this means that the only non-trivial reaction promoted by the interface~$\Ical$ is the heterogeneous conversion of solid phase into carbon gas.
}

\revision{Dividing through by $\rho_e u_{e,i} St_M \,n_i$, the formula above yields the desired $B^\prime$ mass-balance equation
\begin{equation}
\label{eq:bprime}
    y_k^{(f)} - y_{k}^{(e)} + y_{k}^{(f)}B^\prime_{fl} = y_k^{(g)} B^\prime_g + B^\prime_c\delta_{k,k_C},\quad k\in \{1,2\ldots,N_{es}\},
\end{equation}
where $B^\prime_g = \dot{m}^{(g)}/\left(\rho_e u_{e,i} St_M \,n_i\right)$, $B^\prime_c = \chi_{k_C}^\Ical/\left(\rho_e u_{e,i} St_M \,n_i\right)$ and $B^\prime_{fl}$ is defined analogously with $\dot{m}^{(f)}$ on the numerator\footnote{\revision{An anonymous reviewer has kindly pointed out that $B^\prime_f$ is typically used to identify the rate of material removal due to mechanical failure/erosion. 
We therefore use $\Bf$ throughout the paper to refer to the blowing/aspiration rate.}}.
Due to the assumption that the solid phase is converted exclusively into carbon gas, we observe that $\chi_{k_C}^\Ical = -\psi_s^\Ical$, so that, using~\eqref{eq:jump_cond_solid_mass},~$\Bc$ may be expressed as 
\begin{equation}
    \Bc = -\frac{\varepsilon_s \avgrhos v_i n_i}{\rho_e u_{e,i}n_i St_M}.
\end{equation}
(In ablation applications, we have $\Bc \geq 0$, since $v_i n_i \leq 0$ as discussed in section \ref{subsubsec:cons_solid_mass}.)}
For future reference, we also observe that by summing \eqref{eq:bprime} over all $k$ and using the fact that mass fractions sum to $1$, we have $B^\prime_{fl} = B^\prime_g + B^\prime_c$.

\section{$B^\prime$ Formulation from First Principles: Conservation of Energy}
\label{sec:cons_energy}

Here, we follow the same reasoning as in the previous section, and we derive the $B^\prime$ energy-balance equation at a reacting interface.
For this purpose, we consider, once more, the control volume depicted in figure \ref{fig:ctrl_vol}. 

\subsection{Conservation of energy at an ablating surface}

We begin by stating the partial differential equation that governs the conservation of energy on the fluid's side of the interface $\Ical$ (see figure \ref{fig:ctrl_vol}).
As in the previous sections, we consider an ideal gas mixture of $N_s$ species. 
Letting $E = e + (1/2)u_{i}u_{i}$ denote the total (mixture) energy, with $e$ the internal energy, the energy equation on the $(f)$-side of the control volume can be written as 
\begin{equation}
    \label{eq:energy_f}
    \begin{aligned}
    &\frac{\partial}{\partial t}\left(\rho^{(f)}E^{(f)}\right) + \frac{\partial}{\partial x_i}\left(\rho^{(f)}E^{(f)}u_i^{(f)}\right)= \frac{\partial}{\partial x_i}\left( u_j^{(f)}\tau_{i,j}^{(f)}-p^{(f)}u_i^{(f)} + \kappa^{(f)}\frac{\partial T^{(f)}}{\partial x_i} + \mathcal{D}_i^{(f)}\right),
    \end{aligned}
\end{equation}
where 
\begin{equation}
\label{eq:D}
    \mathcal{D}_i^{(f)} = -u_j^{(f)}\sum_{k=1}^{N_s}\rho_k^{(f)}w_{k,i}^{(f)}w_{k,j}^{(f)} - \sum_{k=1}^{N_s}h_k^{(f)}J_{k,i}^{(f)} - \sum_{k=1}^{N_s}\frac{1}{2}w^{(f)}_{k,j}w^{(f)}_{k,j}J_{k,i}^{(f)} + \sum_{k=1}^{N_s}w_{k,j}^{(f)}\tau_{k,i,j}^{(f)}.
\end{equation}
Here, $T$ is the temperature, $\kappa$ is the heat conduction coefficient, $w_{k,i} \coloneqq u_{k,i} - u_i$ is the velocity of species $k$ relative to the mixture velocity, $J_{k,i}^{(f)}$ is defined in \eqref{eq:defn_J}, and $\tau_{i,j} = \sum_{k=1}^{N_s}\tau_{k,i,j}$ is the shear stress tensor.
We refer the reader to \cite{ramshaw2002} for a formal derivation of \eqref{eq:energy_f} for an inviscid ideal gas mixture with zero thermal conductivity. 

On the $(p)$-side of the control volume, occupied by the porous material, the energy equation is often approximated as \citep{chiodi2022}
\begin{equation}
\label{eq:energy_p}
    \frac{\partial}{\partial t}\left(\varepsilon_g\avgrhog
\avgeg + \varepsilon_s\avgrhos\avghs\right) + \frac{\partial}{\partial x_i}\left(\varepsilon_g\avgrhog\avghg\avguig\right) =  \frac{\partial}{\partial x_i}\left(\kappa^{(p)}\frac{\partial\avgT}{\partial x_i}\right).
\end{equation}
Here, $h$ denotes the enthalpy and, as in the previous section, we recall that $\langle\cdot\rangle$ denotes the intrinsic volume average. 
The quantity $\avgT$ is the volume-averaged temperature of the porous material under the assumption of thermal equilibrium between the gaseous phase and the solid phase, and $\kappa^{(p)}$ is the corresponding heat conduction coefficient.
Equation \eqref{eq:energy_p} can be obtained from first principles by volume-averaging the energy equations for the gaseous and solid phases of the porous material.
It should be observed that unclosed terms and several others terms are neglected  during the volume-averaging process, but it is beyond the scope of this paper to provide details on the formal derivation of \eqref{eq:energy_p}.
We refer the reader to, e.g., \cite{whitaker1967} and \cite{gray1976} for details.
A noteworthy observation is that \eqref{eq:energy_p} omits the contribution of the volume-averaged kinetic energy of the gaseous phase (superscript $(g)$) to the total volume-averaged energy of the porous material. 
This has been found to be negligible if the gas exhibits velocities below $100\,m/s$ \citep{martin2008}.



\subsection{The $B^\prime$ energy balance}

We now derive the $B^\prime$ equation for energy conservation across the interface $\Ical$. 
As in section \ref{subsec:bprime_mass_balance}, we stress the fact that the assumptions outlined herein may or may not be physically justified. 
These are made merely to obtain the $B^\prime$ energy equation that is used in existing material response codes. 

Invoking \eqref{eq:jump_cond_varphi} alongside equations \eqref{eq:energy_f} and \eqref{eq:energy_p}, the energy jump condition across the surface~$\Ical$ reads,
\begin{equation}
\label{eq:energy_jump_cond}
    \begin{aligned}    &\left(\rho^{(f)}E^{(f)}\left(u_i^{(f)}-v_i\right)+p^{(f)}u_i^{(f)} - u_j^{(f)}\tau_{i,j}^{(f)} - \kappa^{(f)}\frac{\partial T^{(f)}}{\partial x_i} - \mathcal{D}_i^{(f)}\right)n_i \\
    &- \left(\varepsilon_g\avgrhog\avghg\left(\avguig-v_i\right) - \kappa^{(p)}\frac{\partial\avgT}{\partial x_i} - \varepsilon_s\avgrhos \avghs v_i\right)n_i = \Delta q_{\text{rad}},
    \end{aligned}
\end{equation}
where we recall that $v_i$ denotes the interface velocity and $n_i = n_i^{\Icalp}$ (see figure \ref{fig:ctrl_vol}). 
The term $\Delta q_{\text{rad}}$ denotes the radiative heat transfer at the interface $\Ical$.
This is modelled as an interfacial source term that is analogous in spirit to the term $\psi^\Ical$ in~\eqref{eq:jump_cond_varphi}.

In order to obtain the $B^\prime$ energy balance, we proceed as follows.
Using the fact that $E^{(f)} = h^{(f)} - p^{(f)}/\rho^{(f)} + (1/2)u_i^{(f)}u_i^{(f)}$ and neglecting terms, the first and second terms in the first row of \eqref{eq:energy_jump_cond} become $\rho^{(f)}h^{(f)}\left(u_i^{(f)}-v_i\right)$.
We then neglect $u_j^{(f)}\tau_{i,j}^{(f)}$ and all terms in $\mathcal{D}_i^{(f)}$ (see equation \eqref{eq:D}) except for the second term (i.e., the enthalpy diffusion flux). 
This can be justified using the boundary layer approximation discussed in \cite{eckert1969}.
\revision{Letting $St_H$ denote the heat-transfer Stanton number, and taking $St \coloneqq St_M = St_H$ (i.e., assuming unity Lewis number), we may write}
\begin{equation}
\label{eq:corr_energy}
    -\kappa^{(f)}\frac{\partial T^{(f)}}{\partial x_i} + \sum_{k=1}^{N_s}h_k^{(f)}J_{k,i}^{(f)} = \rho_e u_{e,i} St \left(h^{(f)} - h^{(e)}\right),
\end{equation}
where we recall that superscript/subscript ``e" denotes boundary layer edge quantities.
\revision{While the relationship between unity Lewis number and equal Stanton numbers is well-known and discussed in the literature (see, e.g., \cite{incropera2007,cooper2022}), we present a short derivation in \ref{app:St} to make the manuscript more self-contained.}
Equation \eqref{eq:corr_energy} may be understood as a transfer potential model for heat transfer by convection and diffusion, similar in spirit to the model used to approximate $J_{k,i}$ in \eqref{eq:defn_J}.
Putting this all together, we obtain 
\begin{equation}
    \begin{aligned}
    &\rho^{(f)}h^{(f)}\left(u_i^{(f)}-v_i\right) n_i +\underbrace{\rho_e u_{e,i} St \left(h^{(f)} - h^{(e)}\right) n_i}_{\coloneqq q_{\text{conv}}} \\
    &= \varepsilon_g\avgrhog\avghg\left(\avguig-v_i\right) n_i-\underbrace{\kappa^{(p)}\frac{\partial \avgT}{\partial x_i}n_i}_{\coloneqq q_{\text{cond}}} - \varepsilon_s\avgrhos v_i \avghs n_i + \Delta q_{\text{rad}}.
    \end{aligned}
\end{equation}
This is precisely the energy balance equation displayed, e.g., in \cite{lachaud2014}. 
Dividing through by $\rho_e u_{e,i} St\, n_i$, and recalling the definitions of $\Bg$, $\Bf$ and $\Bc$ in the previous section, the equation above yields the desired $B^\prime$ energy balance
\begin{equation}
\label{eq:bprime_energy}
    h^{(f)}-h^{(e)} + \Bf h^{(f)} = \Bg \avghg - \frac{q_{\text{cond}}}{\rho_e u_{e,i} St\, n_i} + \Bc \avghs + \frac{\Delta q_{\text{rad}}}{\rho_e u_{e,i} St\, n_i}.
\end{equation}
When this equation is solved in practice, the only unknown is $q_{\text{cond}}$, which is then used to specify a Neumann boundary condition on the temperature field $\avgT$.

\section{Extension of the $B^\prime$ Formulation}
\label{sec:extension_bprime}

Despite all the assumptions made in the previous section, the resulting $B^\prime$ formulation should hold for any (positive or negative) values of $\Bg$ and $\Bf$. 
Nonetheless, material response codes and thermodynamics/chemical libraries \citep{lachaud2014,scoggins2020} only consider the case $\Bg \geq 0$. 
Using the control volume in figure \ref{fig:ctrl_vol}, we can see that this corresponds to the case where porous material gases are advected towards the interface $\Ical$ and, by mass conservation, when boundary layer gases are advected away from the interface. 
This scenario is commonly referred to as \emph{blowing}.
However, it is certainly possible that the opposite scenario occurs, where boundary layer gases are advected towards the interface (i.e., \emph{aspiration}) and porous material gases are advected away from the interface.
In this section, we propose a unified $B^\prime$ formulation capable of addressing all these scenarios. 
\revision{Moving forward, mass fractions $y_k$ are to be understood as elemental mass fractions.}

We begin by modifying the transfer potential models used in the original formulation. 
In particular, we write 
\begin{equation}
\label{eq:Jki}
    J_{k,i}^{(f)} = \begin{cases}
        \rho_e u_{e,i}St \left(y_k^{(f)}-y_k^{(e)}\right) & \mathrm{if}\,\, B^\prime_{fl} \geq 0 \\
        \rho_e u_{e,i}St \left(y_k^{(g)}-y_k^{(e)}\right) & \mathrm{if}\,\, B^\prime_{fl} < 0,
    \end{cases}
\end{equation}
and 
\begin{equation}
\label{eq:corr_energy_mod}
    -\kappa^{(f)}\frac{\partial T^{(f)}}{\partial x_i} + \sum_{k=1}^{N_s}h_k^{(f)}J_{k,i}^{(f)}  = 
    \begin{cases}
        \rho_e u_{e,i} St \left(h^{(f)} - h^{(e)}\right) & \text{if } \Bf \geq 0 \\ 
        \rho_e u_{e,i} St \left(h^{(g)} - h^{(e)}\right) & \text{if } \Bf < 0. \\ 
    \end{cases}
\end{equation}
Here, we observe that the need to distinguish between $\Bf \geq 0$ and $\Bf < 0$ in \eqref{eq:Jki} and \eqref{eq:corr_energy_mod} is merely due to notation. 
Specifically, we shall see momentarily that when $\Bf \geq 0$, $y_k^{(f)}$ are the unknown mass fractions that can be computed via Gibbs free energy minimization under the assumption of chemical equilibrium at the wall.
Conversely, when $\Bf < 0$, the equilibrium mass fractions are $y_k^{(g)}$. 
Thus, \eqref{eq:Jki} can be understood as a transfer potential model expressed in terms of the equilibrium mass fractions at the wall. 
This interpretation makes \eqref{eq:Jki} fully consistent with the transfer potential model presented in \cite{eckert1969}.
The same argument holds for the model in \eqref{eq:corr_energy_mod}. 

Given the models \eqref{eq:Jki} and \eqref{eq:corr_energy_mod}, the corresponding $B^\prime$ mass and energy balance equations read 
\begin{empheq}[left = \empheqlbrace]{align}
    &\xi_k - y_{k}^{(e)} + y_{k}^{(f)}B^\prime_{fl} = y_k^{(g)} B^\prime_g + B^\prime_c\delta_{k,k_C},\quad k \in \{1,2,\ldots,N_{es}\}\label{eq:bprime_mass_new}\\
    &\eta - h^{(e)} + \Bf h^{(f)} = \Bg \avghg - \frac{q_{\text{cond}}}{\rho_e u_{e,i} St\, n_i} + \Bc \avghs + \frac{\Delta q_{\text{rad}}}{\rho_e u_{e,i} St\, n_i},
    \label{eq:bprime_energy_new}
\end{empheq}
where 
\begin{equation}
    \xi_k = \begin{cases}
        y_k^{(f)} & \mathrm{if}\,\, B^\prime_{fl} \geq 0 \\
        y_k^{(g)} & \mathrm{if}\,\, B^\prime_{fl} < 0,
    \end{cases},\quad \eta = \begin{cases}
        h^{(f)}  & \text{if } \Bf \geq 0 \\
        h^{(g)} & \text{if } \Bf < 0.
    \end{cases}
\end{equation}
In particular, we see that when $\Bf \geq 0$, equations \eqref{eq:bprime_mass_new} and \eqref{eq:bprime_energy_new} agree with \eqref{eq:bprime} and \eqref{eq:bprime_energy}.
Moreover, we will see that the form of \eqref{eq:bprime_mass_new} and \eqref{eq:bprime_energy_new} (inherited from the transfer potential models in \eqref{eq:Jki} and \eqref{eq:corr_energy_mod}) is such that the unknown equilibrium mass fractions and normalized surface recession rate $\Bc$ are continuous functions of $\Bf$. 
This property provides a well-behaved computational model. 
In the upcoming subsections we discuss the two cases $\Bf \geq 0$ and $\Bf < 0$ in detail.

\subsection{$B^\prime_{fl} \geq 0$ case}

This scenario corresponds to boundary layer gases being advected away from the interface $\Ical$ in figure \ref{fig:ctrl_vol}. 
Recalling that $B^\prime_{fl} = B^\prime_g + B^\prime_c$, we distinguish between two different subcases: $B^\prime_{fl} > B^\prime_c$ and $0 \leq B^\prime_{fl} \leq B^\prime_c$.

\subsubsection{$B^\prime_{fl} > B^\prime_c$}
\label{subsec:case1}

In this case, $B^\prime_g > 0$, meaning that porous material gases are advected towards the interface $\Ical$ in figure \ref{fig:ctrl_vol}.
This is the one and only case considered in the classical $B^\prime$ formulation. 
Here, the unknowns in \eqref{eq:bprime_mass_new} are the mass fractions $y_k^{(f)}$ and~$B^\prime_c$. 
In general, the unknown mass fractions are those associated with the mixture (superscripted either with $(f)$ or $(g)$) that is being advected \emph{away} from the interface.
Since we have less equations than unknowns, solvability is achieved by assuming chemical equilibrium of the species at the interface. 
Under this assumption, the mass fractions~$y_k^{(f)}$ at equilibrium can be computed straightforwardly as a function of pressure, temperature and $B^\prime_g$ via Gibbs free energy minimization \citep{pope2004,scoggins2020}. 
The temperature and pressure are readily available from the boundary conditions, or they can be computed internally by the material response code. 
Likewise, $B^\prime_g$ can be computed internally from $\avguig n_i$ at the surface. 
While it is clear from thermodynamics that the equilibrium composition of a mixture is a function of pressure and temperature, it is helpful to clarify the role of $B^\prime_g$ in this specific application. 

The composition of the equilibrium mixture depends on the initial composition of the reactants. 
In ablation applications, the reactants mixture is assumed to be made up of the ``edge" elemental mixture alongside the porous material gas elemental mixture. 
The elemental composition of the ``edge " mixture can always be expressed in terms of the mass fractions $y_j^{(e)}$ with $j \in \{1,2,\ldots,N_{es}\}$ (e.g., $O$, $H$, $N$ and $C$).  
Similarly, the elemental composition of the mixture on the porous material side can be expressed as $y_j^{(g)}$ with $j \in \{1,2,\ldots,N_{es}\}$.
Thus, the mass fraction of elemental species $j$ in the elemental mixture of reactants is 
\begin{equation}
\label{eq:initial_yj}
    y_j^{\text{reactants}} = \frac{y_j^{(e)} + B^\prime_g y_j^{(g)}}{\sum_{k=1}^{N_{es}}\left(y_k^{(e)} + B^\prime_g y_k^{(g)}\right)}.
\end{equation}
Clearly, different values of $B^\prime_g$ lead to different reactants mixtures, and it is therefore clear that the resulting equilibrium mixture will also be a function of $B^\prime_g$. 

Once the equilibrium mass fractions $y_k^{(f)}$ are obtained,~$B^\prime_c$ can be obtained directly from \eqref{eq:bprime} using $B^\prime_{fl} = B^\prime_g + B^\prime_c$. 
In particular, fixing $k = k_C$, (where we recall that $k_C$ is the index pointing to monatomic carbon~$C$), we have
\begin{equation}
\label{eq:bc_1}
    B^\prime_c = \frac{y_{k}^{(g)}-y_k^{(f)}}{y_k^{(f)}-1}B^\prime_g + \frac{y_{k}^{(e)} - y_k^{(f)}}{y_k^{(f)}-1}.
\end{equation}
The process just described is usually tabulated (i.e., precomputed) as a function of pressure, temperature and $B^\prime_g$. 
Hence the name $B^\prime$ table. 
In the energy equation \eqref{eq:bprime_energy_new}, the only unknown is $q_{\text{cond}}$, which sets a Neumann boundary condition for the the temperature field, $h^{(f)}$ is taken to be the enthalpy of the wall equilibrium mixture (given by the $B^\prime$ table), and $\langle h\rangle^{(g)}$ is taken to be the enthalpy associated with the elemental composition on the porous material side of the interface.

\subsubsection{$0\leq B^\prime_{fl} \leq B^\prime_c$}
\label{subsec:case2}

In this case, $B^\prime_g \leq 0$, meaning that porous material gases are advected away from the interface $\Ical$. 
The unknowns in \eqref{eq:bprime_mass_new} are $\Bc$ as well as $y_k^{(f)}$ and $ y_k^{(g)}$, since both boundary layer gases and porous material gases are being advected away from the interface.  
Since chemical equilibrium calculations yield one equilibrium mixture, it is clear that $y_k^{(f)} = y_k^{(g)}$.
It is worth remarking that while in the previous case the equilibrium mixture was a function of pressure, temperature and $B^\prime_g$, here the mixture is only a function of pressure and temperature. 
In fact, since porous material gases are advected away from the interface, the mass fractions of the elemental mixture of reactants are given by the elemental ``edge" composition alone,
\begin{equation}
    y_j^{\text{reactants}} = \frac{y_j^{(e)}}{\sum_{k=1}^{N_{es}} y_k^{(e)}}.
\end{equation}
Since the reactants mixture does not depend on $\Bg$, the equilibrium mixture will also be independent of $\Bg$. 
Once the equilibrium mass fractions $y_k^{(f)} = y_k^{(g)}$ are computed, \eqref{eq:bprime_mass_new} gives us (with $k=k_C$)
\begin{equation}
\label{eq:bc_2}
    \Bc = \frac{y_k^{(e)} - y_k^{(f)}}{y_k^{(f)}-1} = \frac{y_k^{(e)} - y_k^{(g)}}{y_k^{(g)}-1}.
\end{equation}
This equation is quite interesting, as it states that in this regime $\Bc$ is independent of $\Bg$. 
We can also readily check that if we evaluate \eqref{eq:bc_1} at $\Bg = 0$, this agrees with~\eqref{eq:bc_2}, meaning that $\Bc$ is continuous at $\Bf = \Bc$. 
In the energy equation \eqref{eq:bprime_energy_new}, $h^{(f)} = \langle h\rangle^{(g)}$ and they are taken to be equal to the enthalpy of the wall equilibrium mixture computed using the $B^\prime$ table.

\subsection{$\Bf < 0$ case}
\label{subsec:case3}

We now consider the case $\Bf < 0$, which corresponds to boundary layer gases being advected towards the interface $\Ical$ in figure \ref{fig:ctrl_vol}. 
Thus, the unknowns in \eqref{eq:bprime_mass_new} are~$y_k^{(g)}$ and $\Bc$. 
The boundary layer mass fractions $y_k^{(f)}$, on the other hand, are set equal to the edge mass fractions $y_k^{(e)}$.
This is equivalent to assuming a frozen boundary layer, where the ``edge" elemental composition is equal to the elemental composition in close proximity of the wall. 
By setting $y_k^{(f)} = y_k^{(e)}$, the equilibrium mixture becomes independent of $\Bg$, and thus only a function of pressure and temperature. 
This can be seen immediately once we observe that the reactants mixture is defined by equation \eqref{eq:initial_yj} with $y_k^{(g)}$ replaced by $y_k^{(f)}$ and $\Bg$ replaced by $\Bf$. 
Given the equilibrium mass fractions $y_k^{(g)}$, formula \eqref{eq:bprime_mass_new} can be solved for $\Bc$ with $k=k_C$,
\begin{equation}
\label{eq:bc_3}
    \Bc = \frac{y_k^{(g)}-y_k^{(f)}}{y_k^{(f)}-1}\Bg + \frac{y_k^{(e)} - y_k^{(g)}}{y_k^{(g)}-1}.
\end{equation}
First, we observe that since the equilibrium mass fractions are independent of $\Bg$, then $\Bc$ is a linear function of $\Bg$. 
Second, if we evaluate \eqref{eq:bc_3} at $\Bf = 0$ (i.e., $\Bg = -\Bc$) we can see after some manipulation that this agrees with \eqref{eq:bc_2}.
Thus, $\Bc$ is continuous at $\Bf = 0$, as desired.
In the energy equation \eqref{eq:bprime_energy_new}, $h^{(f)} = c_p T$, where $T$ is the wall temperature and $c_p$ is the specific heat capacity based on ``edge" quantities (due to the fact that we take $y_k^{(f)}=y_k^{(e)}$), and $\langle h \rangle^{(g)}$ is taken to be the enthalpy of the wall equilibrium mixture delivered by the $B^\prime$ table.

\begin{figure}
    \centering
    \begin{tikzpicture}
    \begin{axis}[axis lines=middle,axis equal,xlabel={$B^\prime_g$},ylabel={$B^\prime_{fl}$},x label style={at={(axis description cs:1.05,0.06)}}]
    \addplot[black,samples=100] coordinates{(-4,-1.5) (-1,0)};
    \addplot[black,samples=100] coordinates{(-1,0) (0,1)};
    \addplot[black,samples=100,domain=0:2] {1.5*x + 1 + 0.5*x^2};
    \draw [decorate, decoration={brace,amplitude=5pt,raise=2pt}] (0,1) -- (0,0) node [midway, xshift=-1mm, yshift=2mm, auto, outer sep=10pt,font=\tiny]{$B^\prime_{c,0} = \frac{y_k^{(e)}-y_k^{(f)}}{y_k^{(f)}-1}$};
    \draw [decorate, decoration={brace,amplitude=5pt,raise=2pt}] (0,0) -- (-1,0) node [midway, yshift=1mm, xshift=0mm, auto, outer sep=10pt,font=\tiny]{$B^\prime_{c,0}$};
    \end{axis}
    \end{tikzpicture}
    \caption{Schematic (not to scale) of the three $B^\prime_{fl}$ regimes discussed in section \ref{sec:extension_bprime}. Recall that $B^\prime_{fl} = \Bg + \Bc$. For $\Bf > \Bc$, the curve is nonlinear due to the nonlinear dependence of $\Bc$ on $\Bg$ (equation \eqref{eq:bc_1}). For $0\leq \Bf\leq \Bc$, the curve is linear with slope $1$, since $\Bc$ does not depend on $\Bg$ (equation \eqref{eq:bc_2}). Finally, for $\Bf< 0$, the curve is linear with slope given by \eqref{eq:bc_3}.
    }
    \label{fig:schematic_bprime}
\end{figure}
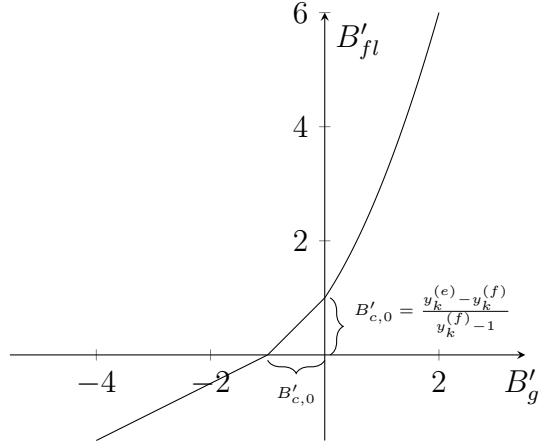

We conclude this section by pointing the reader's attention to figure \ref{fig:schematic_bprime}, which shows a schematic of the three $\Bf$ regimes discussed thus far.
This shows that if we account for the inflow of gases into the porous material (i.e., $\Bg < 0$), $\Bc$ will always be greater than or equal to $B^\prime_{c,0}$, which is the value at $\Bg = 0$.
In particular, if $\Bg < 0$ is small enough that $0 \leq \Bf\leq \Bc$, then by equation \eqref{eq:bc_2} we see that $\Bc = B^\prime_{c,0}$.
Since $\Bc$ is directly proportional to the recession velocity of the interface $\Ical$, this implies that the classical $B^\prime$ formulation will predict a recession velocity that is exactly equal to the recession velocity predicted by the new $B^\prime$ formulation. 
However, if $\Bg < 0$ is large enough that $\Bf < 0$, then by \eqref{eq:bc_3} $\Bc > B^\prime_{c,0}$, and the classical framework will predict a recession velocity that is lower than that predicted by the new formulation.

\revision{Finally, in order to facilitate the implementation of the new $B^\prime$ formulation in existing material response codes, we provide some representative pseudocode in Algorithm \ref{alg:bprime}.
Given a modified $B^\prime$ table (which can be easily generated following the guidelines in \ref{app:tables} or using the scripts in \url{https://github.com/albertopadovan/Modified_Bprime}), 
the algorithm shows that existing material response codes that are already equipped to use the classical $B^\prime$ formulation should require very little additional logical to handle the extended $B^\prime$ formulation.
}


\begin{algorithm}
\caption{Algorithmic outline of the new $B^\prime$ formulation }\label{alg:bprime}
\begin{algorithmic}[1]
\Require Modified $B^\prime$ table generated following \ref{app:tables} and/or the scripts in the repository \url{https://github.com/albertopadovan/Modified_Bprime}, pressure~$p$ and temperature $\langle T\rangle $ at the ablating surface, and gas velocity $\langle u_i\rangle^{(g)} n_i$ normal to the ablating surface.
\Ensure Normalized ablation rate $\Bc$, enthalpy $h_w$ of the equilibrium mixture at the surface, and solution to equations \eqref{eq:bprime_mass_new} and \eqref{eq:bprime_energy_new}.

\vspace{2ex}
\State Compute $\Bg$ (see definition in section \ref{subsec:bprime_mass_balance}) using $\langle u_i\rangle^{(g)} n_i$.
\State Using $\Bg$, $p$ and $\langle T \rangle$, compute $\Bc$ and $h_w$ using the modified $B^\prime$ table (which is constructed so that \eqref{eq:bprime_mass_new} is automatically satisfied with the appropriate $\xi_k$).
\State Compute $\Bf = \Bg + \Bc$.
\If{$\Bf > \Bc$}
    \State Solve \eqref{eq:bprime_energy_new} with $\Bf \geq 0$, $h^{(f)} = h_w$ and $\langle h\rangle^{(g)}$ taken to be the formation enthaply of the elemental composition on the porous material side of the interface (see section \ref{subsec:case1}).
\ElsIf{$0 \leq \Bf \leq \Bc$}
    \State Solve \eqref{eq:bprime_energy_new} with $\Bf \geq 0$ and $h^{(f)} = \langle h \rangle^{(g)} = h_w$ (see section \ref{subsec:case2}).
\Else 
    \State Solve \eqref{eq:bprime_energy_new} with $\Bf < 0$, $h^{(f)} = h^{(e)}$ and $\langle h \rangle^{(g)} = h_w$ (see section \ref{subsec:case3}).
\EndIf
\end{algorithmic}
\end{algorithm}

\subsection{A note on the blowing/suction correction}

When we are interested in computing the material response of a porous material to an external flow, but we are not resolving (or computing) the response of the fluid to the material dynamics, the Stanton number $St$ is usually corrected to account for the effect of a non-zero velocity (i.e., suction/blowing) at the interface. 
In particular, given the Stanton number $St_0$ associated with no suction or blowing, the corrected Stanton number $St$ is given by 
\begin{equation}
    \label{eq:blowing_correction}
    \frac{St}{St_0} = \frac{\log\left(1 + 2\lambda \Bf\right)}{2\lambda \Bf},
\end{equation}
where $\lambda > 0$.
This correction was initially derived from the incompressible (laminar) velocity boundary layer equations to correct the skin friction coefficient in the presence of suction or blowing \citep{kays1993}. 
Given that the thermal and concentration boundary layer equations with unity Prandtl and Lewis numbers are analogous to the velocity boundary layer equations \citep{eckert1969,incropera2007}, it follows immediately that, under the same assumptions, the same correction can be used to correct the Stanton number. 
The derivation in \cite{kays1993} for laminar incompressible boundary layers led to $\lambda = 0.5$. 
According to \cite{moyer1968}, $\lambda=0.4$ has been reported to be better suited for turbulent flows.

Since the derivation in \cite{kays1993} holds for any positive and negative non-zero velocities at the surface (i.e., positive and negative $\Bf$, in our case), the correction in \eqref{eq:blowing_correction} may be used for both positive and negative values of $\Bf$. 
The only caveat is that \eqref{eq:blowing_correction} requires $2\lambda\Bf > -1$, otherwise the logarithm is not defined. 
This simply means that as $2\lambda \Bf$ approaches $-1$ from the right, the assumptions that originally led to \eqref{eq:blowing_correction} no longer hold. 
We remark that \eqref{eq:blowing_correction} is well-posed for $\Bf = 0$, since
\begin{equation}
    \lim_{2\lambda\Bf\to 0}\,\, \frac{St}{St_0} = 1.
\end{equation}
In practical applications, it is possible for $2\lambda\Bf$ to be less than or equal to~$-1$, in which case use of \eqref{eq:blowing_correction} would lead to computational issues. 
We resolve the issue by artificially lower bounding $2\lambda\Bf$ to $-0.9$. 
We close this section by observing that while the blowing/suction correction is the most popular approach to account for suction and blowing in a boundary layer, a few authors \citep{demuelenaere2012,cooper2023} have proposed formulations that bypass the need to correct the Stanton number using \eqref{eq:blowing_correction}.

\section{Application to a TACOT Wedge}
\label{sec:results}

In this section we compare the new $B^\prime$ formulation with the classical $B^\prime$ formulation on a two-dimensional pyrolyzing and ablating TACOT \citep{tacot2018} wedge,
whose geometry is shown in figure \ref{fig:wedge}a.

\subsection{Description of the computational setup}

\revisiontwo{The ``Theoretical Ablative Composite for Open Testing" (TACOT) is a porous material consisting of two solid phases (non-reacting fibers and a reacting matrix), with a virgin (i.e., non-pyrolyzed) solid volume fraction of $0.20$ and a charred (i.e., pyrolyzed) solid volume fraction of $0.15$.}
The response of the material to a prescribed boundary condition (described below) is simulated using the in-house material response solver CHyPS, whose governing equations and computational discretization are described in section III of \cite{chiodi2022}.
\revisiontwo{In particular, all conservation laws are obtained via volume averaging, with the conservation of gaseous mass and solid mass taking the form of equations \eqref{eq:cons_mass_k_volavg} and \eqref{eq:cons_mass_solid_avg}, respectively. 
The volumetric source terms $\langle \psi_k\rangle^{(g)}$ and $\langle \psi_s\rangle^{(s)}$ enter the formulation due to the heterogenous conversion of solid mass to gaseous mass promoted by pyrolysis.
Pyrolysis itself is modelled via three chemical reactions with Arrenhius coefficients specified in table~1 of \cite{chiodi2022}.
Conservation of momentum within the porous material is reduced to Darcy's law, while conservation of energy (which takes the form of \eqref{eq:energy_p}) is posed under the assumption of thermal equilibrium.
Finally, the mesh movement induced by ablation is handled with the Arbitrary Lagrangian Eulerian (ALE) formulation.
}

\revisiontwo{The treatment of the gas and solid properties inside the TACOT wedge are discussed in detail in sections III\,D and IV of \cite{chiodi2022}.
In particular, gas properties are assumed to be functions of pressure and temperature only, while solid properties are assumed to be functions of temperature and of the pyrolysis progress variable (denoted $\tau$ in the notation of \cite{chiodi2022}, with $\tau = 0$ indicating the virgin state and $\tau = 1$ the charred state).
Gas and solid properties, as well as bulk properties (e.g., thermal conductivity and permeability) are determined via the TACOT lookup tables available in \citep{tacot2018}.
Moreover, TACOT is treated as an isotropic porous material and the volumetric gas composition is held constant at $y_O = 0.115$, $y_C = 0.206$ and $y_H = 0.679$ according to the TACOT model \citep{tacot2018}.
It is worth observing that a more advanced volumetric gas chemistry model could be used, and it could include species tracking and equilibrium/non-equilibrium chemistry. 
In that case, a chemistry boundary condition can be provided by the new $B^\prime$ formulation when gas is advecting into the material.
The new $B^\prime$ formulation is implemented according to algorithm \ref{alg:bprime}. 
The classical $B^\prime$ formulation is implemented analogously, except that $\Bg$ is artificially set to $0$ in step 1 of algorithm \ref{alg:bprime} when boundary layer gases enter the porous material.
Finally, the radiation term $\Delta q_{\text{rad}}$ in equation \eqref{eq:bprime_energy_new} is modelled following the Stefan-Boltzmann law for a grey body.   
}

\begin{figure}
\centering
\begin{minipage}{0.48\textwidth}
\begin{tikzonimage}[trim= 85 15 20 10,clip,width=0.83\textwidth]{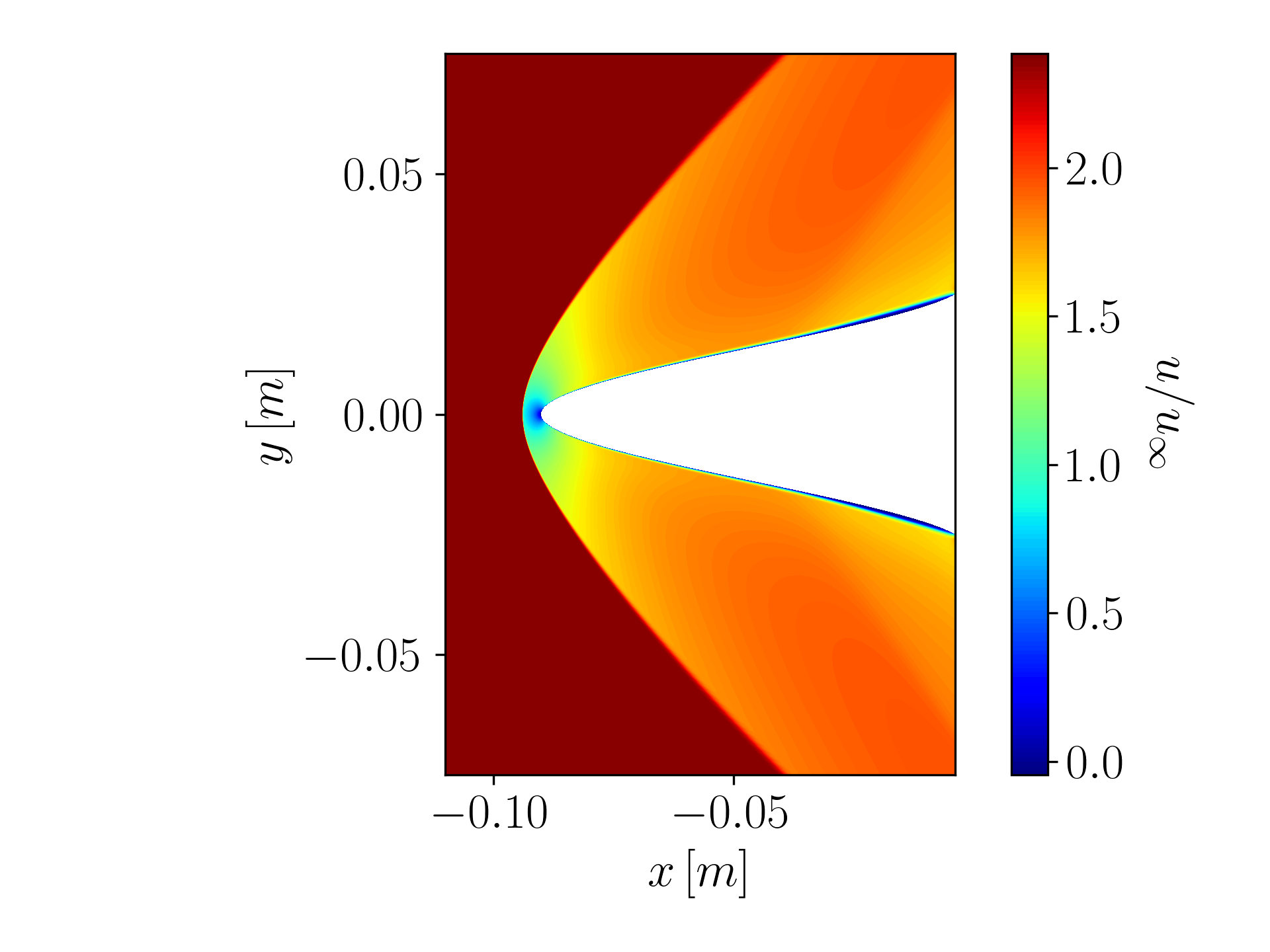}
\node at (0.67,0.9) {\small $\textit{(a)}$};
\end{tikzonimage}
\end{minipage}
\hspace{-5ex}
\begin{minipage}{0.48\textwidth}
\begin{tikzonimage}[trim= 15 15 5 10,clip,width=1.03\textwidth]{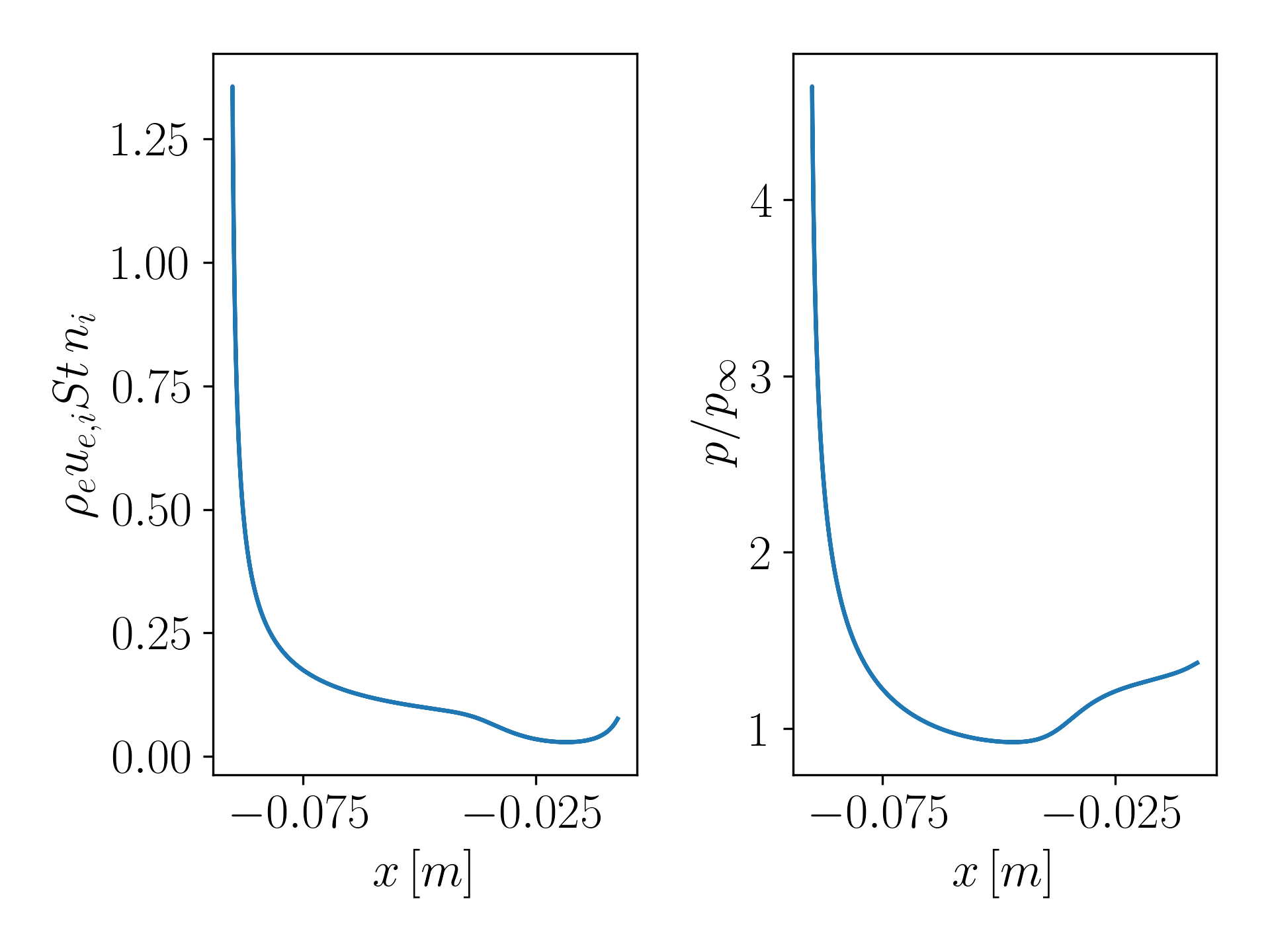}
\node at (0.4,0.9) {\small $\textit{(b)}$};
\node at (0.9,0.9) {\small $\textit{(c)}$};
\end{tikzonimage}
\end{minipage}
\caption{\textit{(a)} Steady-state normalized streamwise velocity field around the wedge, \textit{(b)} $\rho_e u_{e,i}St \,n_i$ profile and \textit{(c)} normalized pressure profile at the wedge surface. Here, $p_{\infty}=10\,kPa$.
The wedge surface at $t = 0$ is parameterized according to the equation $x = c_1 y^3 + c_2 y^2 + c_3$, where $c_1 = -7966.80539304$, $c_2 = 336.99725483$ and $c_3 = -0.09014195$.
}
\label{fig:wedge}
\end{figure}

The dynamics of the material are fully specified by the pressure and normalized heat flux profiles on the surface of the wedge. 
Nominal normalized pressure $p/p_\infty$ and normalized heat flux $\rho_{e}u_{e,i}St\,n_i$ profiles, shown in figures \ref{fig:wedge}b and \ref{fig:wedge}c, are obtained from the steady-state solution of a Mach-$2$ flow around the wedge. 
In particular, \revision{we used the in-house solver PlasCom2} to solve the compressible Navier-Stokes equations at freestream conditions $M_\infty=2$, $T_\infty=1000 \mathrm{K}$, $p_\infty=10\mathrm{kPa}$, and Reynolds number $Re_\infty = 1.1 \times 10^6$ based on a freestream characteristic length $L=1$. 
The fluid was modeled as a single-species ideal gas with $\gamma=1.4$ and $R=287\,\mathrm{J}/\left(\mathrm{kg-K}\right)$.
Viscosity and thermal conductivity were modeled with a viscous power law of $\mu = \mu_{298}\left(T/T_{298}\right)^{0.666}$ and $Pr=0.72$. 
The material interface boundary condition was enforced with a no-slip, impermeable, isothermal wall at $1000\,K$, with non-zero pressure gradient.
\revision{The $\rho_{e}u_{e,i}St\,n_i$ profiles were computed using equation \eqref{eq:corr_energy} with $h^{(e)}=h_\infty\left(1 + \left(\sqrt{Pr}\left(\gamma-1\right)/2\right)M_\infty^2 \right)$.}
The $B^\prime$ tables were generated using Mutation++ \citep{scoggins2020} with the NASA-9 thermodynamics database, and assuming an ``edge" elemental composition $y_N^{(e)} = 0.790$, $y_O^{(e)} = 0.210$, and a pyrolysis gas elemental composition $y_O^{(\text{pyro})} = 0.115$, $y_C^{(\text{pyro})} = 0.206$ and $y_H^{(\text{pyro})} = 0.679$.
Details are described in \ref{app:tables}.
\revision{It is also important to remark that, throughout, ablation is treated exclusively as a surface phenomenon and volume ablation is neglected.}

\begin{table}
\centering
\begin{tabularx}{0.8\textwidth} { 
  | >{\centering\arraybackslash}X 
  | >{\centering\arraybackslash}X 
  | >{\centering\arraybackslash}X | }
 \hline
  & $\alpha$ & $p_\infty$ \\
 \hline
 Case 1  & $1$  &  $10\,kPa$ \\
 \hline
 Case 2  & $1$  &  $1\,kPa$ \\
 \hline
 Case 3  & $1/4$  &  $1\,kPa$ \\
 \hline
 Case 4  & $1/4$  &  $10\,kPa$ \\
\hline
\end{tabularx}
\caption{Scaling factors $\alpha$ for the normalized heat flux boundary condition, and external reference pressure $p_\infty$.}
\label{tab:cases}
\end{table}

In order to study how the two $B^\prime$ formulations behave under different heating and external pressure conditions, we run four different simulations.  
In particular, we specify the normalized heat flux boundary condition as 
\begin{equation}
    \alpha \rho_{e}u_{e,i}St\,n_i,
\end{equation}
where $\alpha$ is a scaling factor, and we vary the external reference pressure $p_\infty$. 
The values of $\alpha$ and $p_\infty$ for the four different cases are listed in table \ref{tab:cases}.
The material response code is initialized with zero heat flux and uniform pressure $p_{\infty}$ on the wedge surface at $t=0$, and it is brought (via linear interpolation) to the desired surface boundary condition over a ramping period of $0.01\,s$. 
After that, we observe the response of the wedge for a total of $0.5\,s$.
\revision{For all cases considered herein, we will see that the length of the temporal interval $t\in [0,0.5]\,s$ is sufficient for initial transients to decay and to observe post-transient dynamics.}
Throughout, we use $\lambda = 0.5$ in the blowing correction \eqref{eq:blowing_correction}.

\subsection{Discussion of the results}
\begin{figure}
\centering
\begin{minipage}{0.48\textwidth}
\begin{tikzonimage}[trim= 15 15 5 10,clip,width=0.98\textwidth]{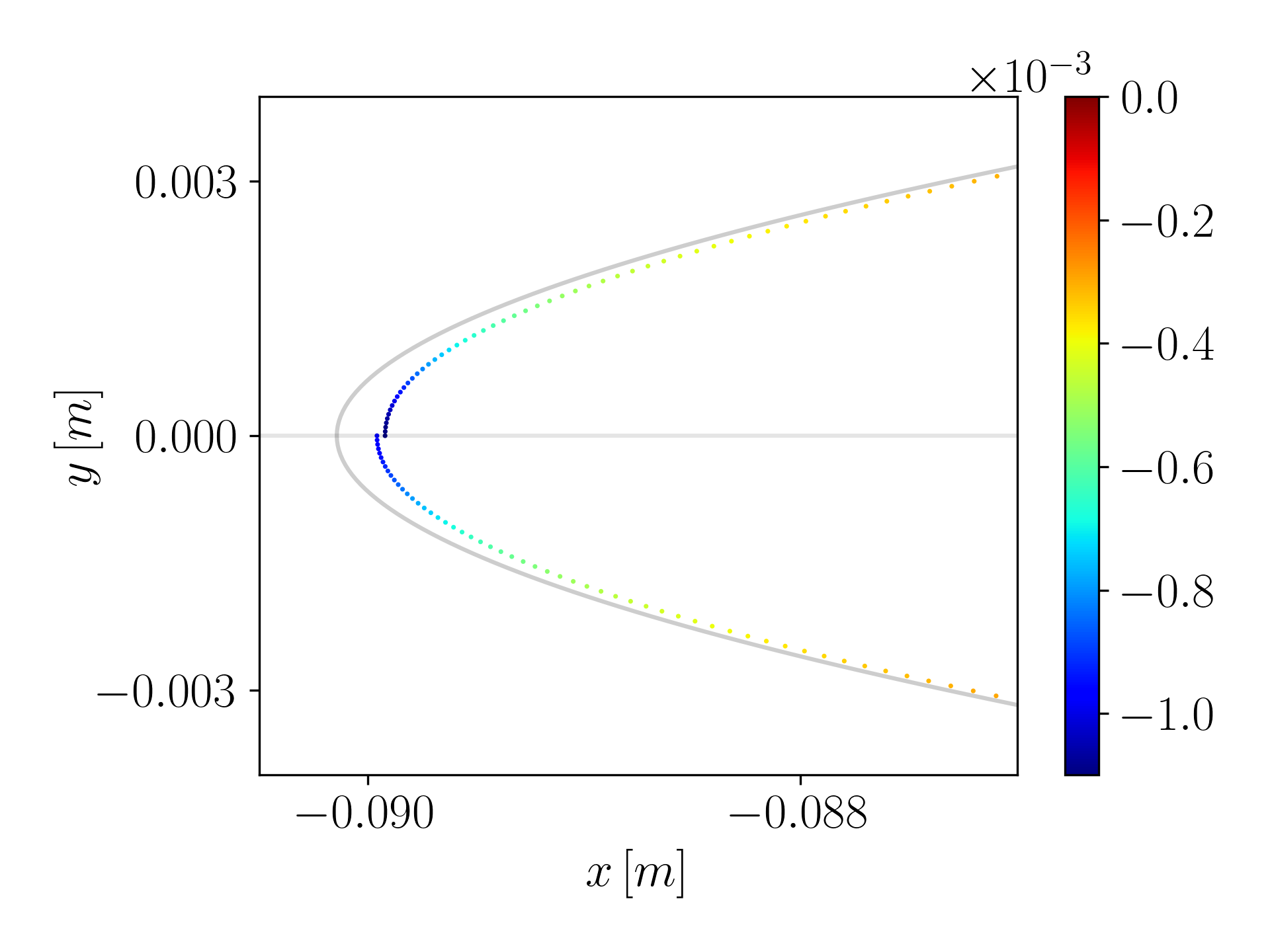}
\node at (0.23,0.22) {\small $\textit{(a)}$};
\node at (0.30,0.86) {\small Case 1};
\end{tikzonimage}
\end{minipage}
\begin{minipage}{0.48\textwidth}
\begin{tikzonimage}[trim= 15 15 5 10,clip,width=0.98\textwidth]{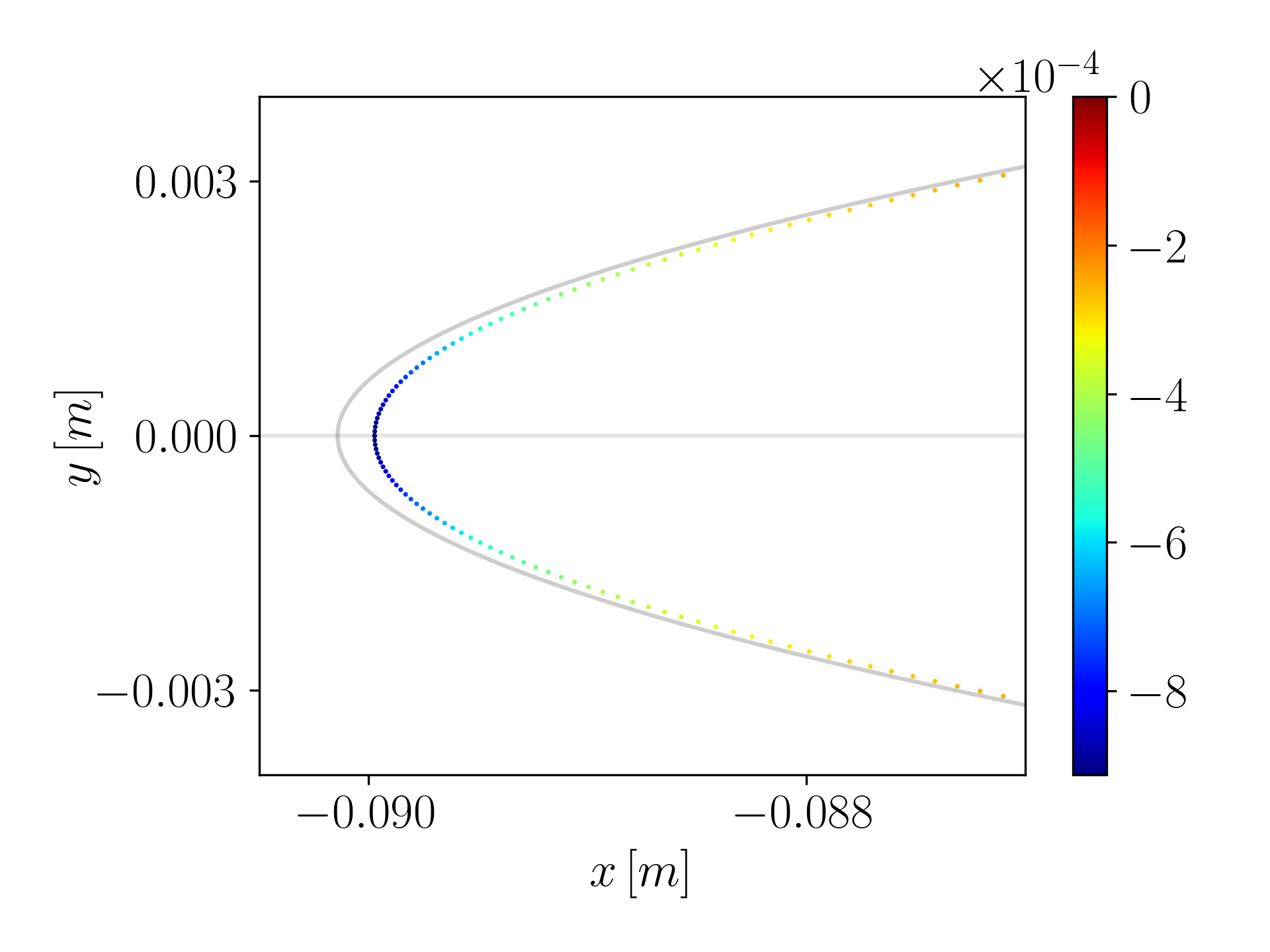}
\node at (0.23,0.22) {\small $\textit{(b)}$};
\node at (0.30,0.86) {\small Case 2};
\end{tikzonimage}
\end{minipage}
\begin{minipage}{0.48\textwidth}
\begin{tikzonimage}[trim= 15 15 5 10,clip,width=0.98\textwidth]{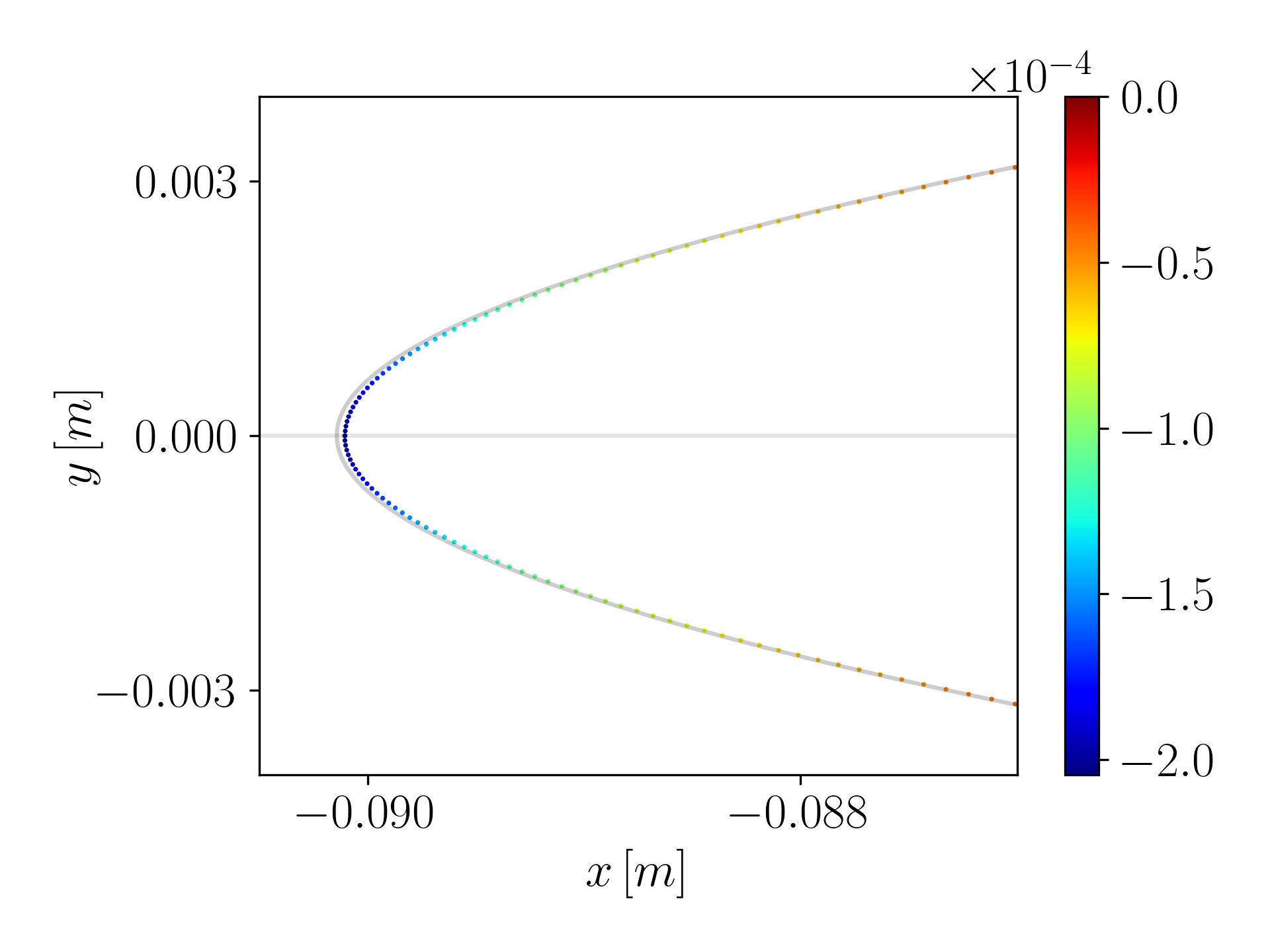}
\node at (0.23,0.22) {\small $\textit{(c)}$};
\node at (0.30,0.86) {\small Case 3};
\end{tikzonimage}
\end{minipage}
\begin{minipage}{0.48\textwidth}
\begin{tikzonimage}[trim= 15 15 5 10,clip,width=0.98\textwidth]{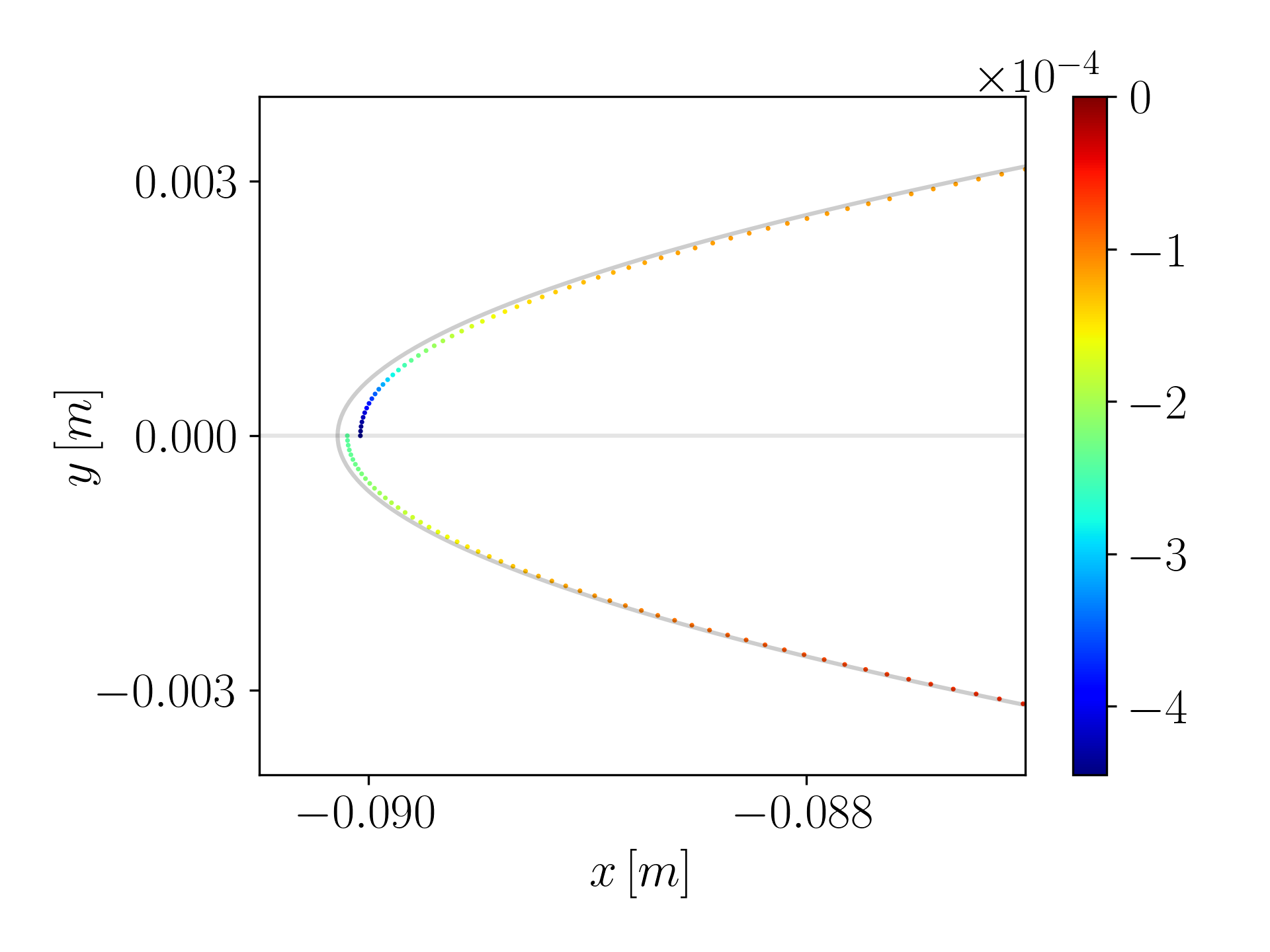}
\node at (0.23,0.22) {\small $\textit{(d)}$};
\node at (0.30,0.86) {\small Case 4};
\end{tikzonimage}
\end{minipage}
\caption{Wedge geometry at  time $t = 0.20\,s$, colorcoded by the instantaneous surface recession velocity \revision{(in meters per second)} normal to the surface. The top half ($y \geq 0$) is the prediction using the new $B^\prime$ formulation, the bottom half is the prediction using the classical $B^\prime$ formulation. The gray line shows the geometry at $t = 0$.}
\label{fig:recession_uncoupled_t020}
\end{figure}

\begin{figure}
\centering
\begin{minipage}{0.48\textwidth}
\begin{tikzonimage}[trim= 15 15 5 10,clip,width=0.98\textwidth]{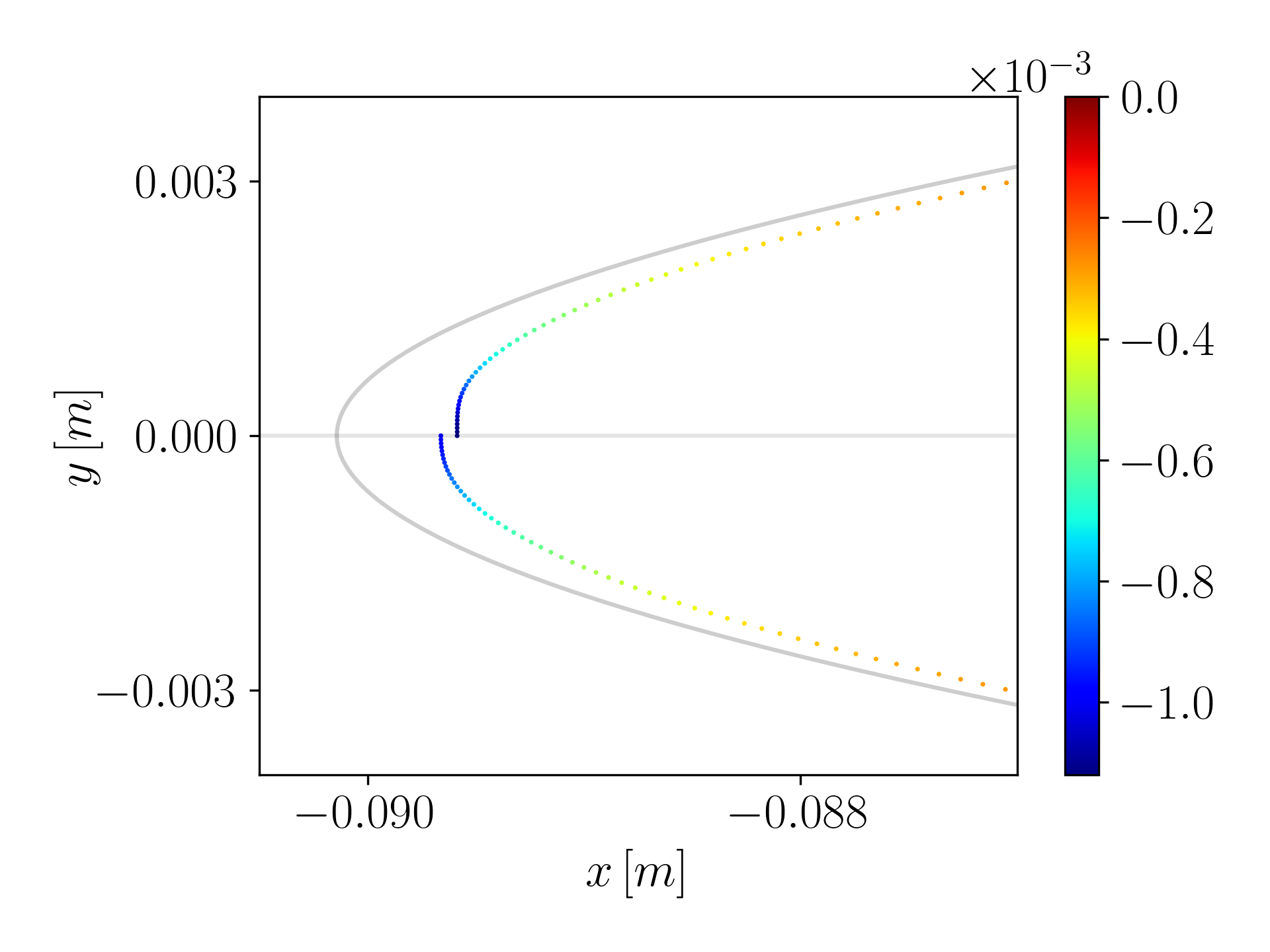}
\node at (0.23,0.22) {\small $\textit{(a)}$};
\node at (0.30,0.86) {\small Case 1};
\end{tikzonimage}
\end{minipage}
\begin{minipage}{0.48\textwidth}
\begin{tikzonimage}[trim= 15 15 5 10,clip,width=0.98\textwidth]{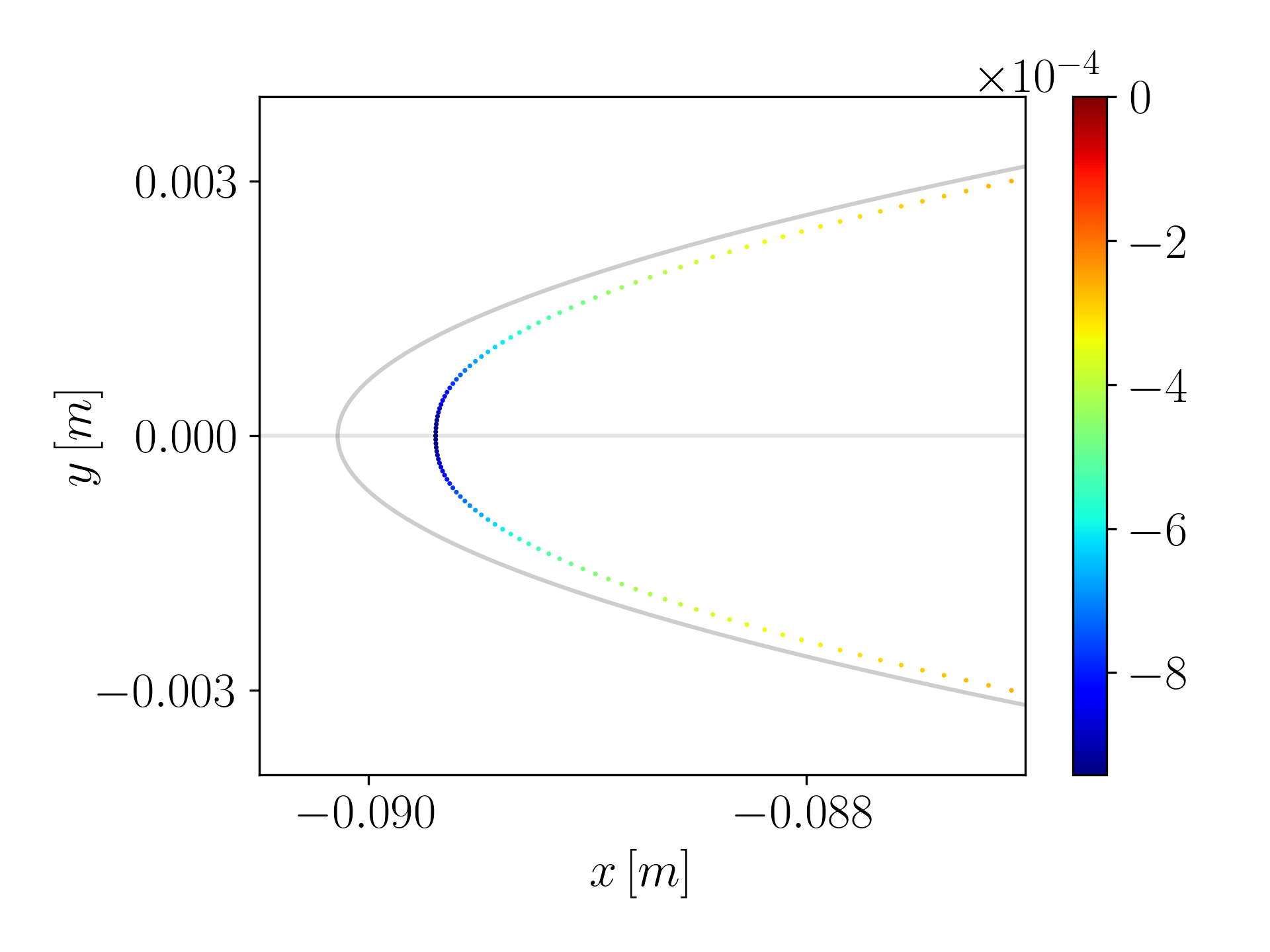}
\node at (0.23,0.22) {\small $\textit{(b)}$};
\node at (0.30,0.86) {\small Case 2};
\end{tikzonimage}
\end{minipage}
\begin{minipage}{0.48\textwidth}
\begin{tikzonimage}[trim= 15 15 5 10,clip,width=0.98\textwidth]{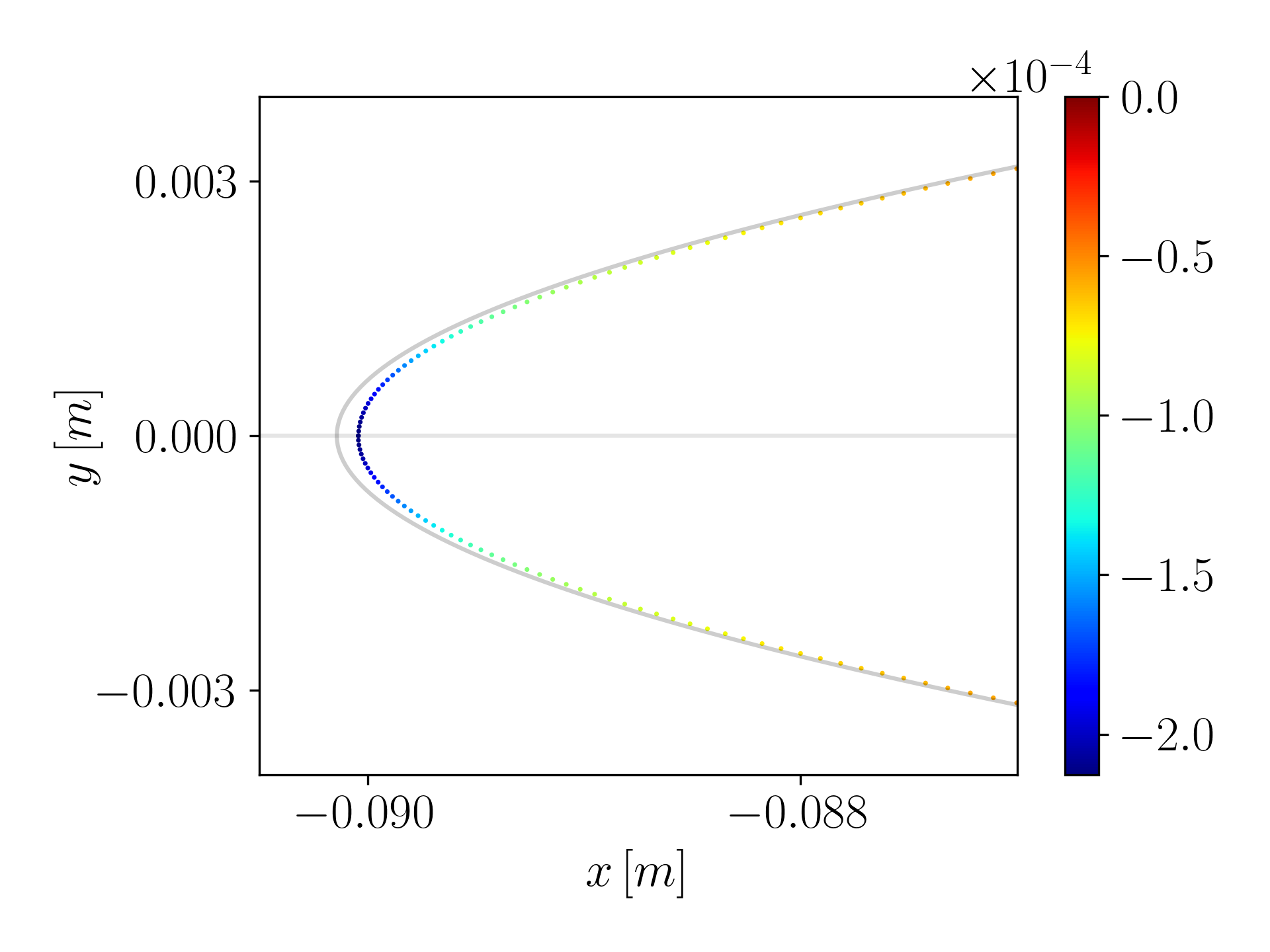}
\node at (0.23,0.22) {\small $\textit{(c)}$};
\node at (0.30,0.86) {\small Case 3};
\end{tikzonimage}
\end{minipage}
\begin{minipage}{0.48\textwidth}
\begin{tikzonimage}[trim= 15 15 5 10,clip,width=0.98\textwidth]{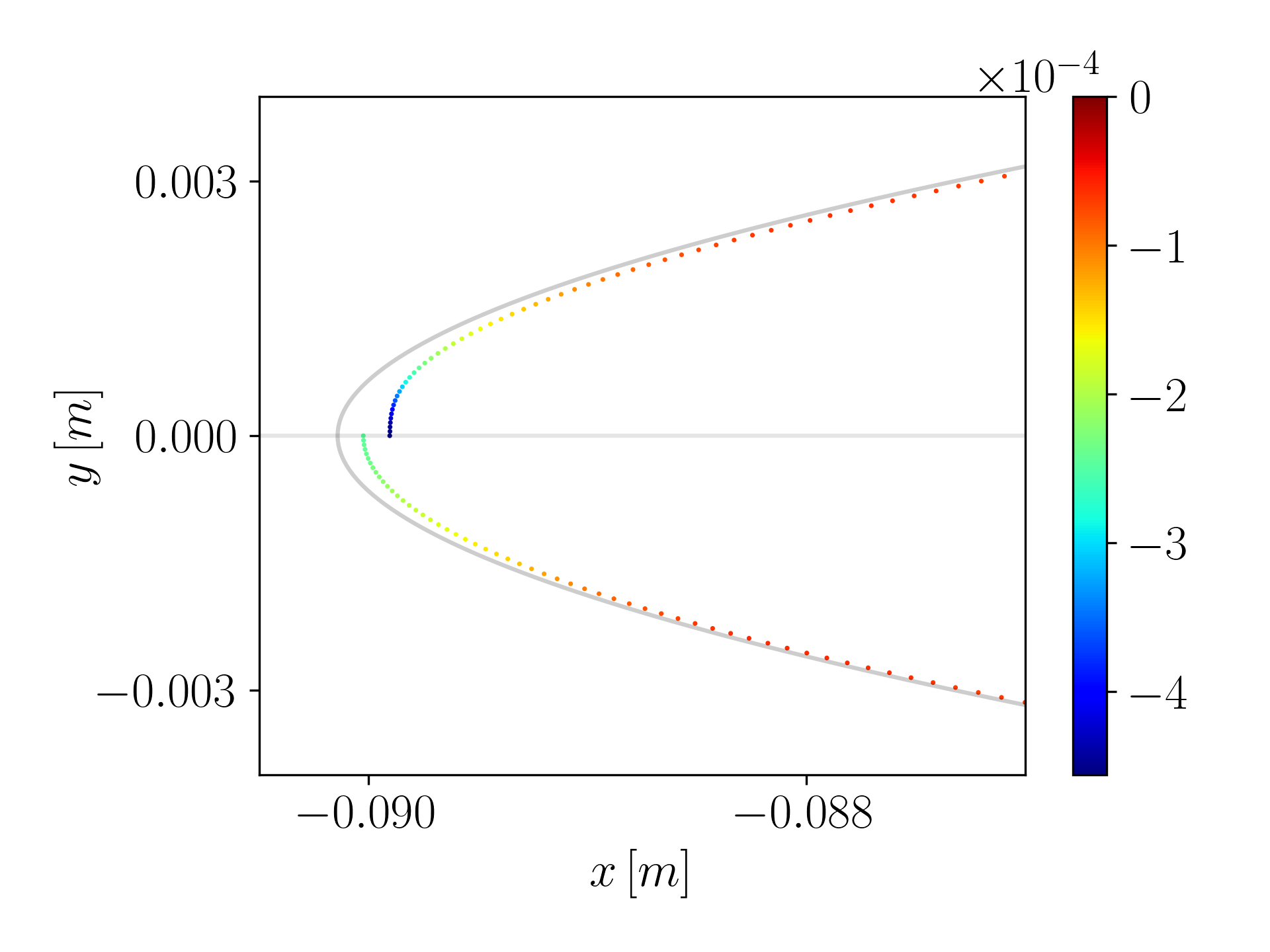}
\node at (0.23,0.22) {\small $\textit{(d)}$};
\node at (0.30,0.86) {\small Case 4};
\end{tikzonimage}
\end{minipage}
\caption{Analog of figure \ref{fig:recession_uncoupled_t020} at time $t = 0.50\, s$.}
\label{fig:recession_uncoupled_t050}
\end{figure}

\begin{figure}
\centering
\begin{minipage}{0.48\textwidth}
\begin{tikzonimage}[trim= 15 100 5 90,clip,width=0.95\textwidth]{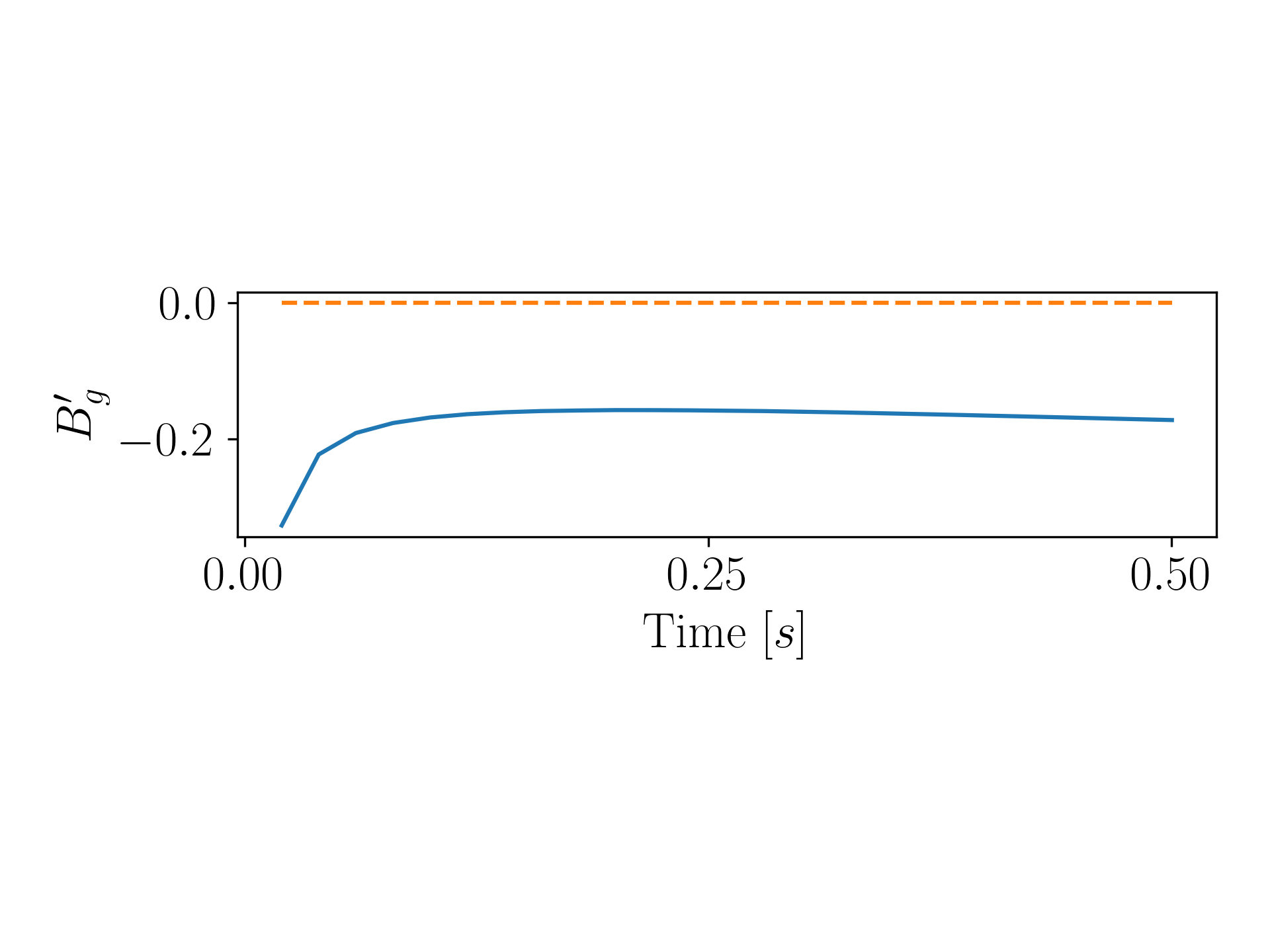}
\node at (0.9,0.45) {\small $\textit{(a)}$};
\end{tikzonimage}
\end{minipage}
\begin{minipage}{0.48\textwidth}
\begin{tikzonimage}[trim= 15 100 5 90,clip,width=0.95\textwidth]{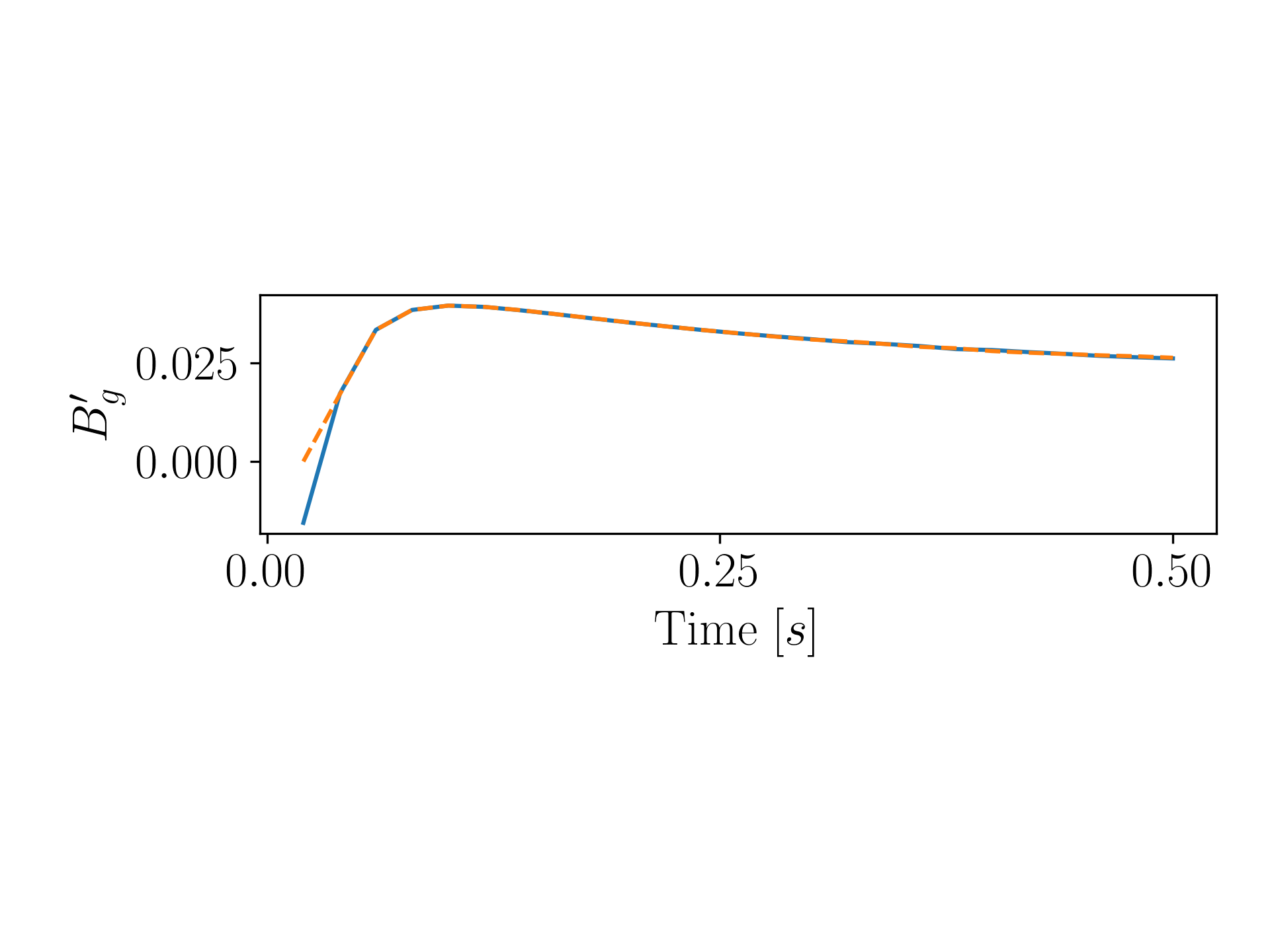}
\node at (0.9,0.45) {\small $\textit{(b)}$};
\end{tikzonimage}
\end{minipage}
\begin{minipage}{0.48\textwidth}
\begin{tikzonimage}[trim= 15 100 5 90,clip,width=0.95\textwidth]{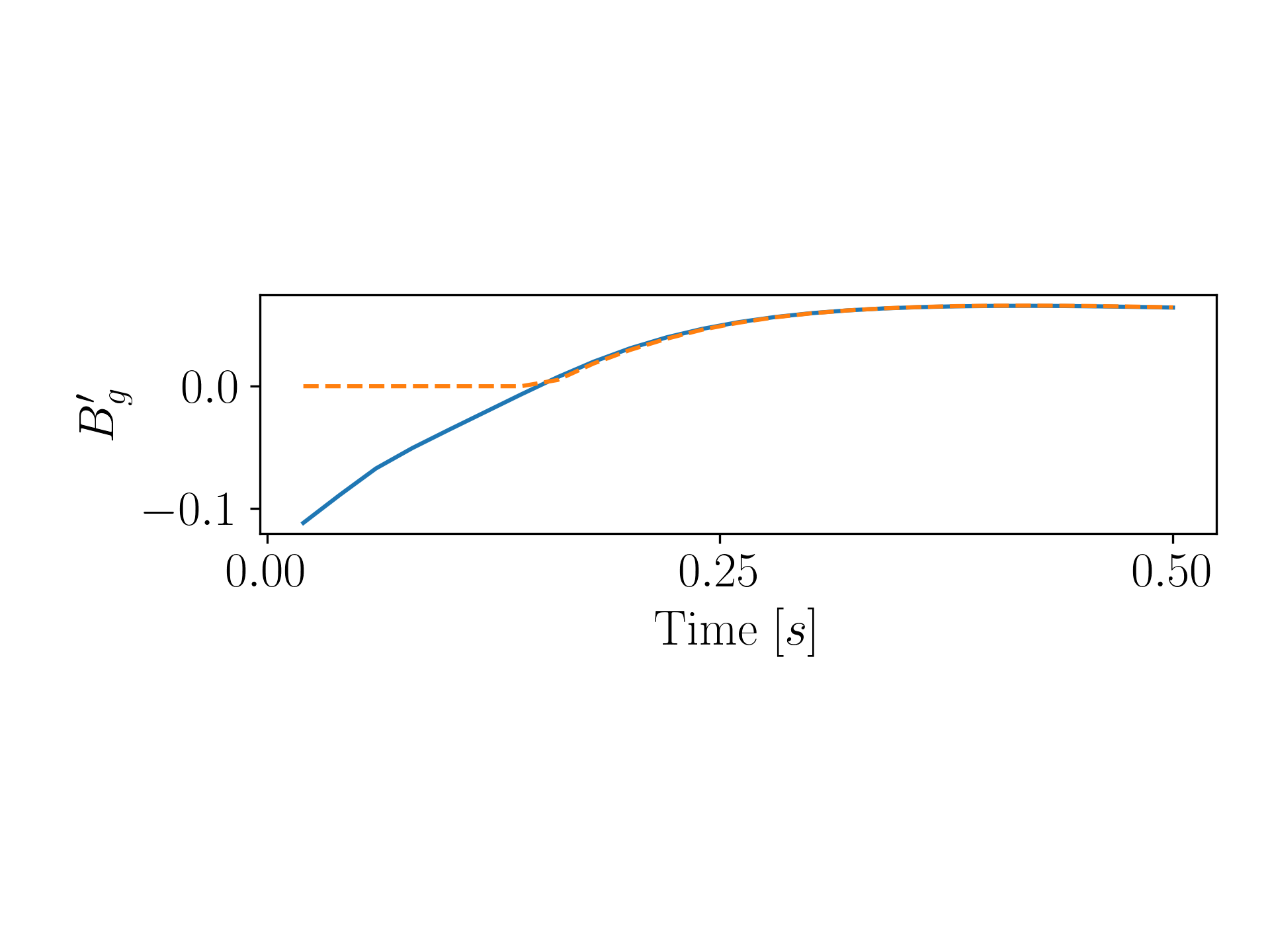}
\node at (0.9,0.45) {\small $\textit{(c)}$};
\end{tikzonimage}
\end{minipage}
\begin{minipage}{0.48\textwidth}
\begin{tikzonimage}[trim= 15 100 5 90,clip,width=0.95\textwidth]{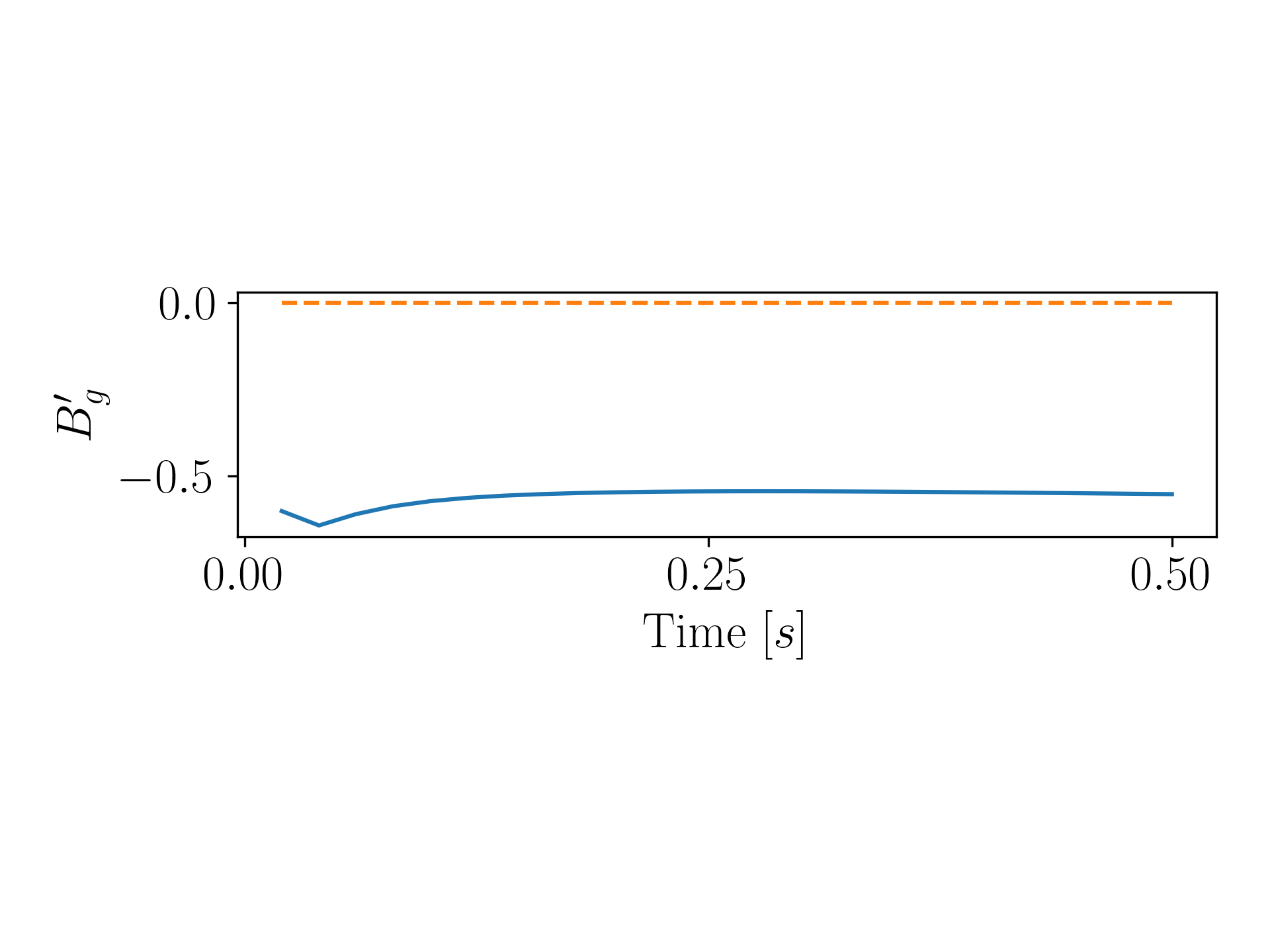}
\node at (0.9,0.45) {\small $\textit{(d)}$};
\end{tikzonimage}
\end{minipage}
\caption{Time history of $\Bg$ at the wedge leading edge for cases 1-4 (panels $(a)$-$(d)$). The solid line corresponds to the new $B^\prime$ formulation, while the dashed line to the classical $B^\prime$ formulation.}
\label{fig:bg_uncoupled}
\end{figure}

\begin{figure}
\centering
\begin{minipage}{0.48\textwidth}
\begin{tikzonimage}[trim= 15 100 5 90,clip,width=0.95\textwidth]{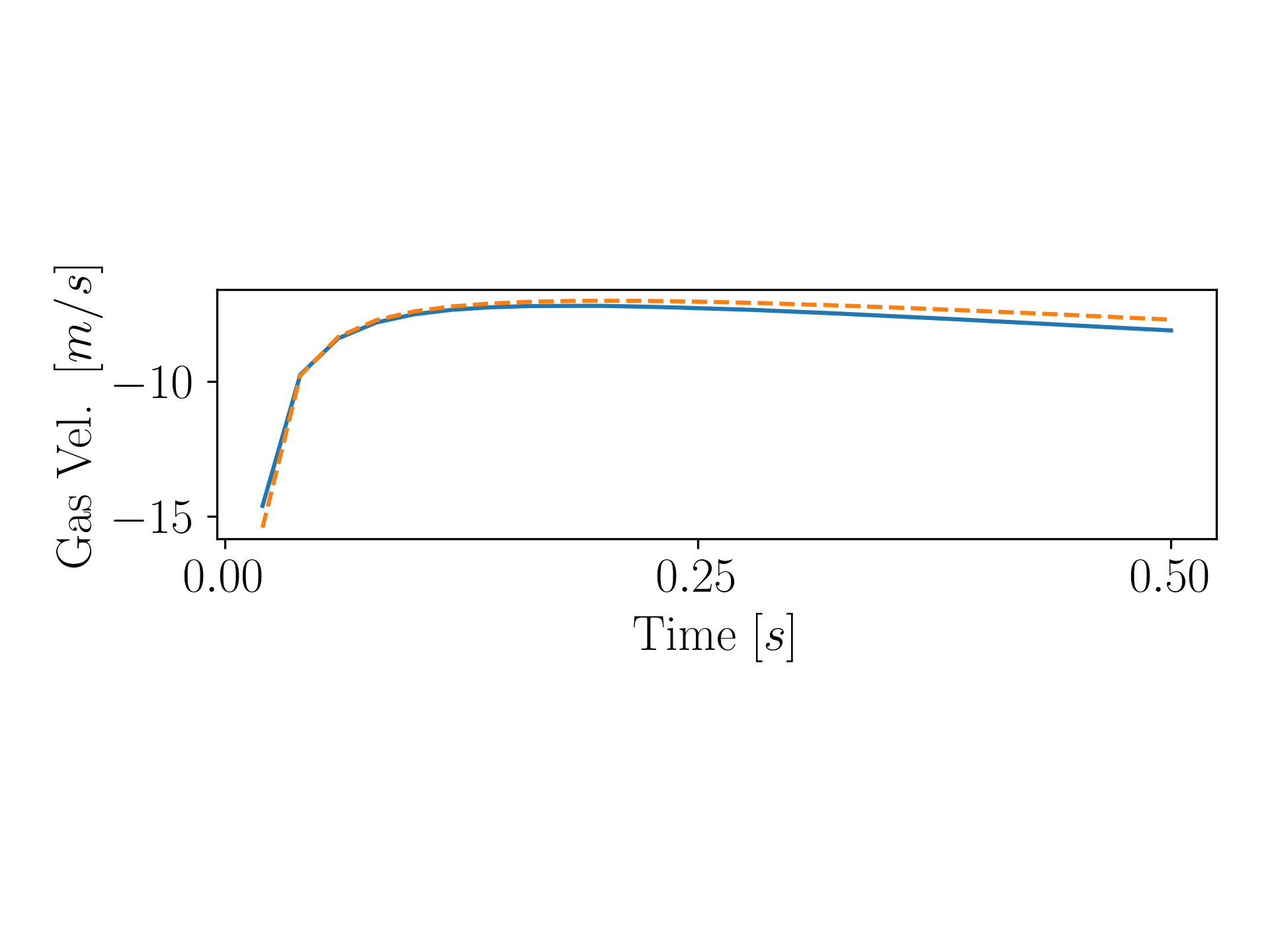}
\node at (0.9,0.45) {\small $\textit{(a)}$};
\end{tikzonimage}
\end{minipage}
\begin{minipage}{0.48\textwidth}
\begin{tikzonimage}[trim= 10 100 5 90,clip,width=0.95\textwidth]{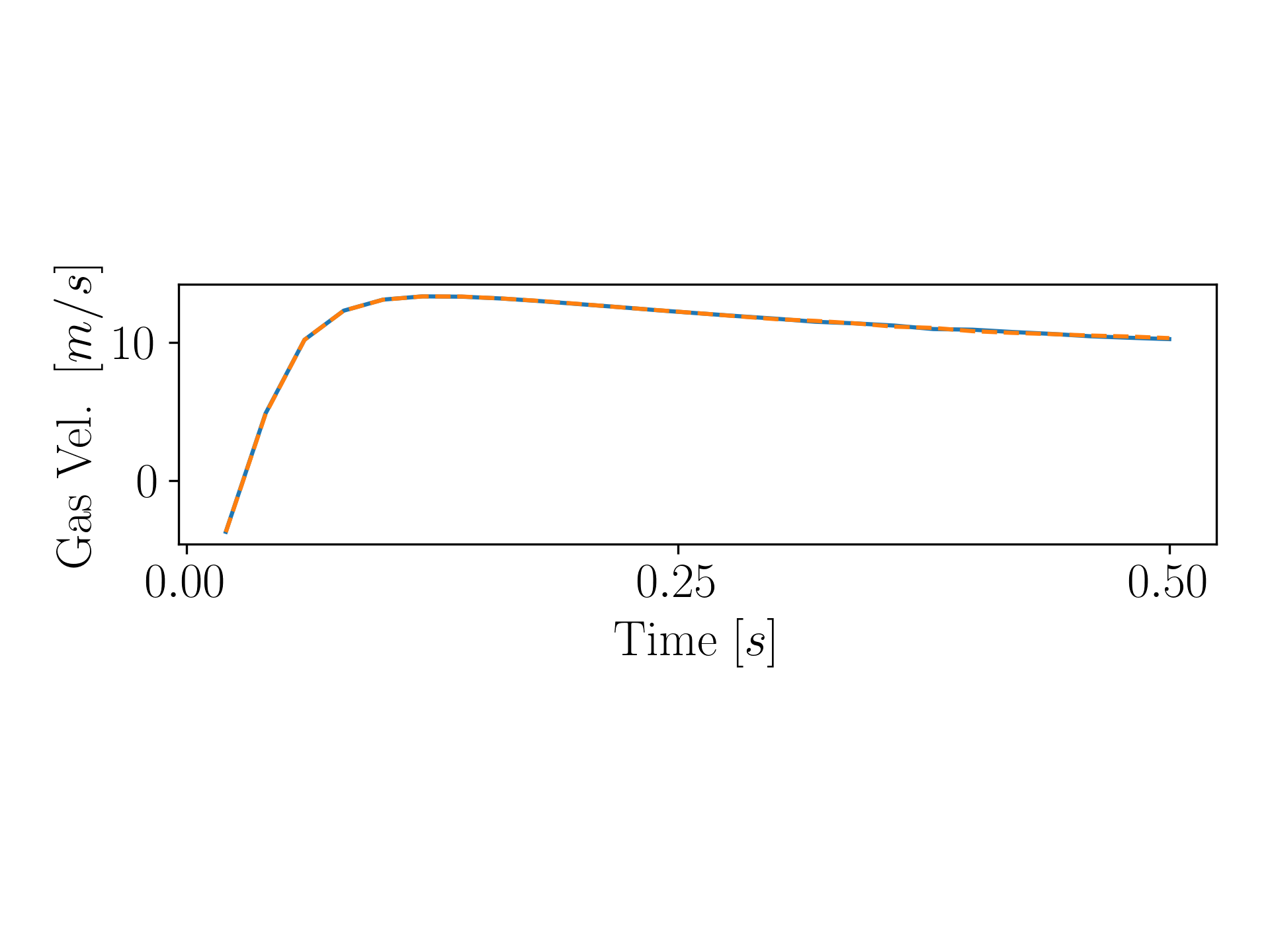}
\node at (0.9,0.45) {\small $\textit{(b)}$};
\end{tikzonimage}
\end{minipage}
\begin{minipage}{0.48\textwidth}
\begin{tikzonimage}[trim= 5 100 5 90,clip,width=0.95\textwidth]{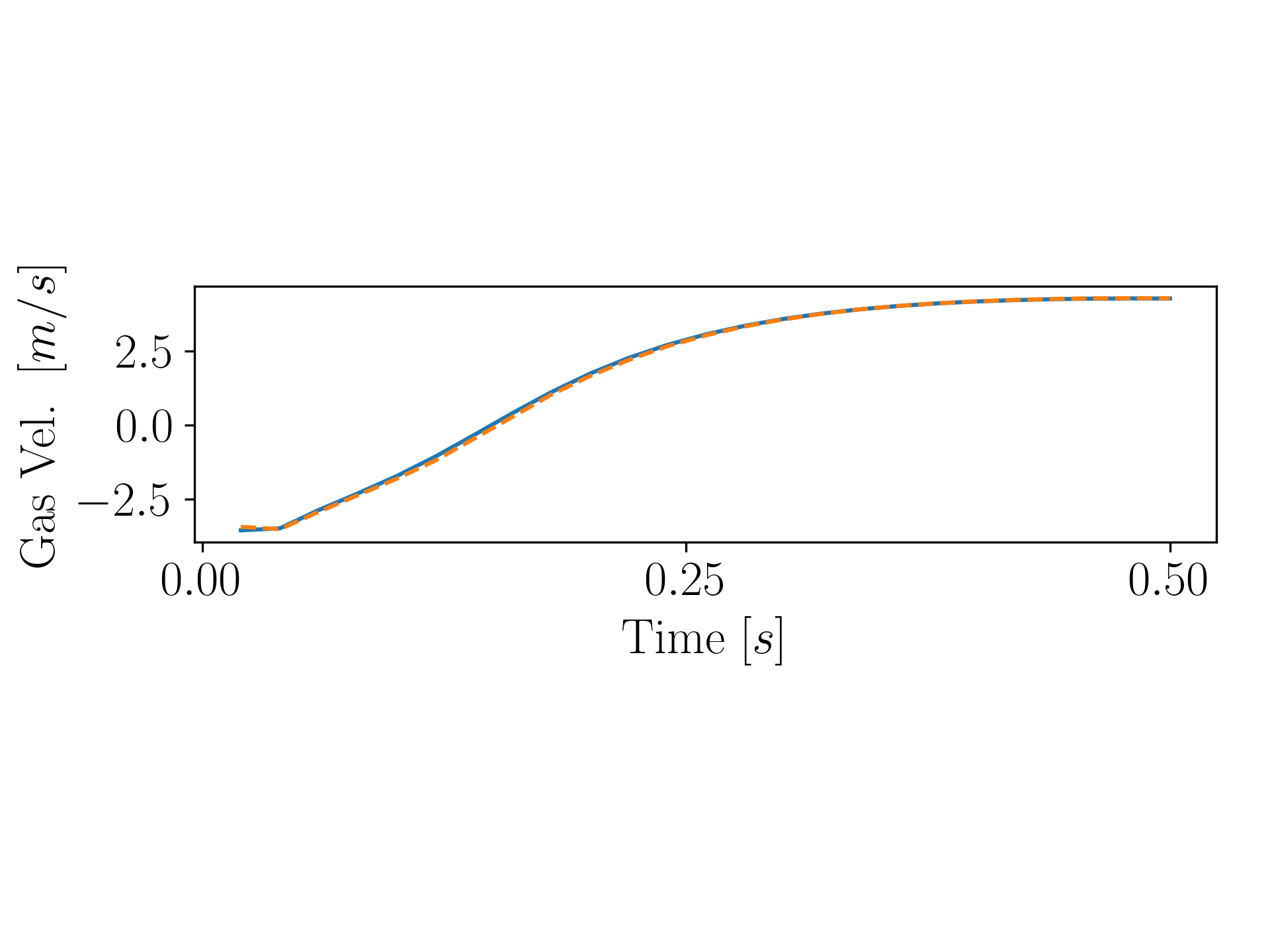}
\node at (0.9,0.45) {\small $\textit{(c)}$};
\end{tikzonimage}
\end{minipage}
\begin{minipage}{0.48\textwidth}
\begin{tikzonimage}[trim= 15 100 5 90,clip,width=0.95\textwidth]{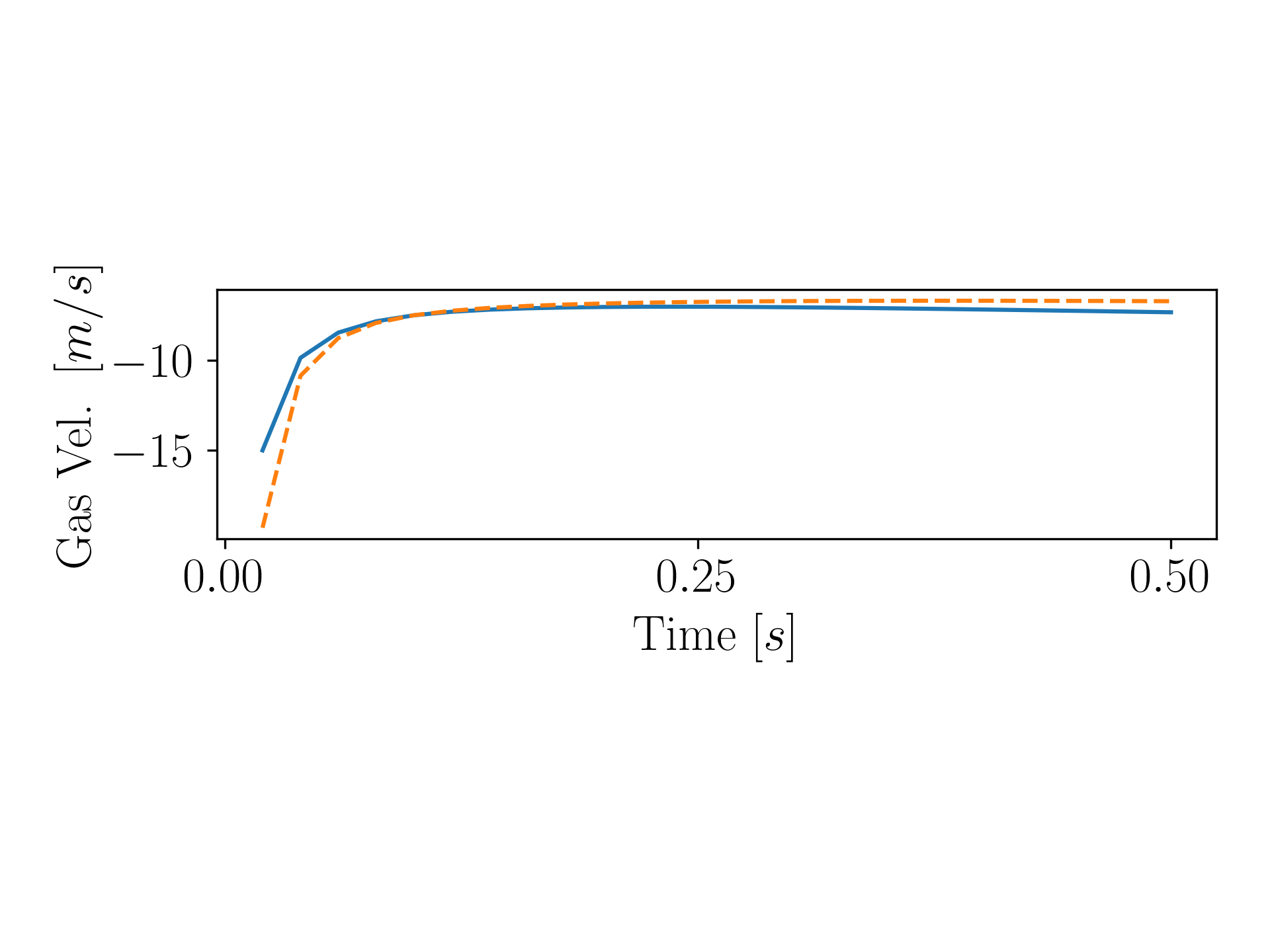}
\node at (0.9,0.45) {\small $\textit{(d)}$};
\end{tikzonimage}
\end{minipage}
\caption{Analog of figure \ref{fig:bg_uncoupled}, for the time-history of the gas velocity normal to the surface at the wedge leading edge. (Negative values indicate that gases are entering the porous material.)}
\label{fig:gas_vel_uncoupled}
\end{figure}

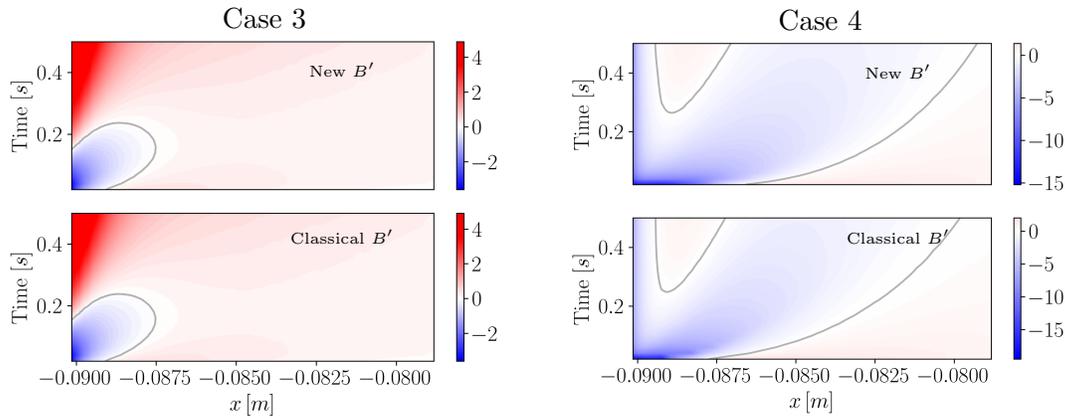
\begin{figure}
\centering
\begin{minipage}{0.48\textwidth}
\begin{tikzonimage}[trim= 15 10 20 10,clip,width=0.95\textwidth]{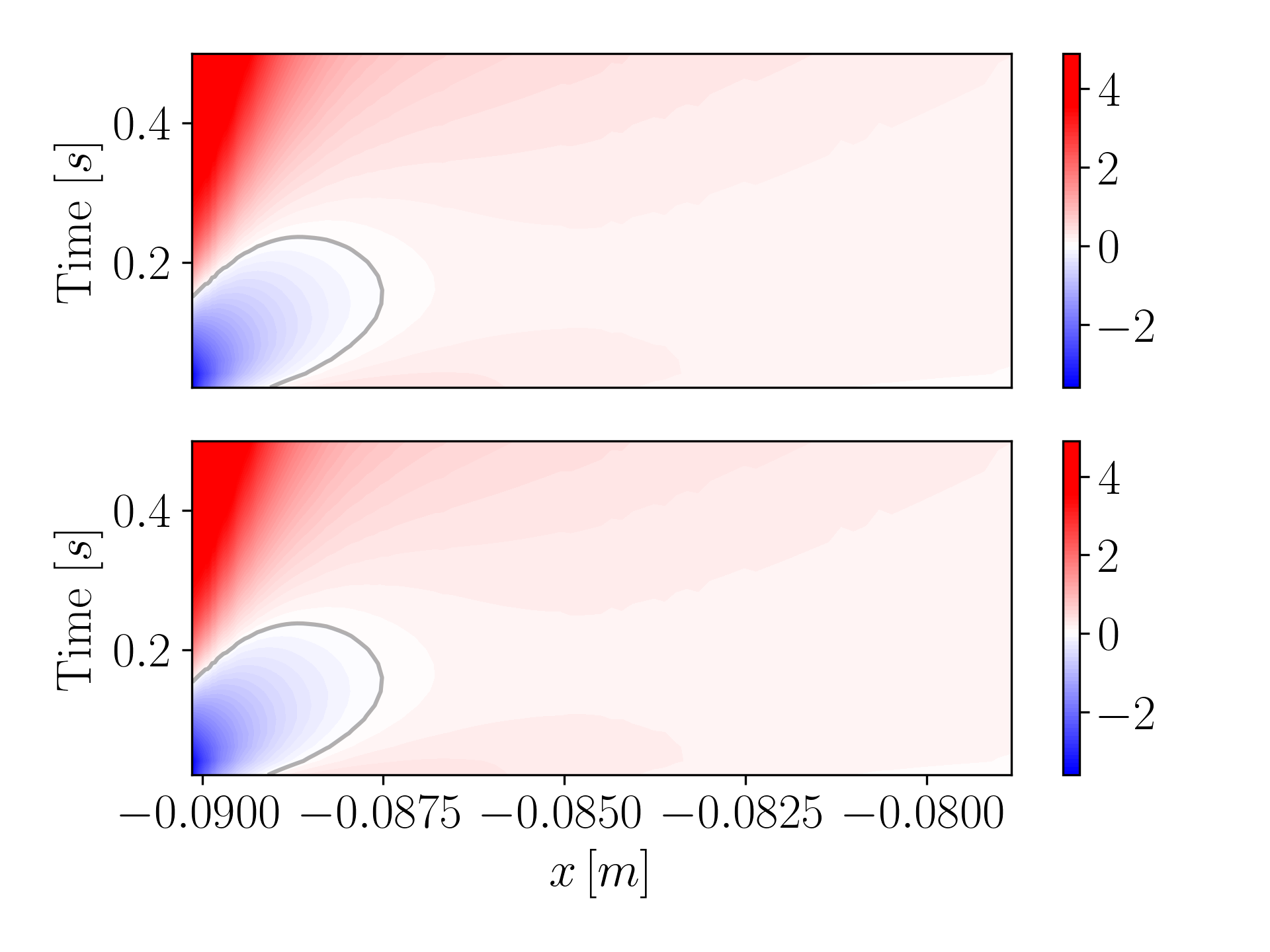}
\node at (0.5,1.03) {\small Case 3};
\node at (0.65,0.9) {\tiny New $B^\prime$};
\node at (0.65,0.48) {\tiny Classical $B^\prime$};
\end{tikzonimage}
\end{minipage}
\begin{minipage}{0.48\textwidth}
\begin{tikzonimage}[trim= 10 10 20 10,clip,width=0.95\textwidth]{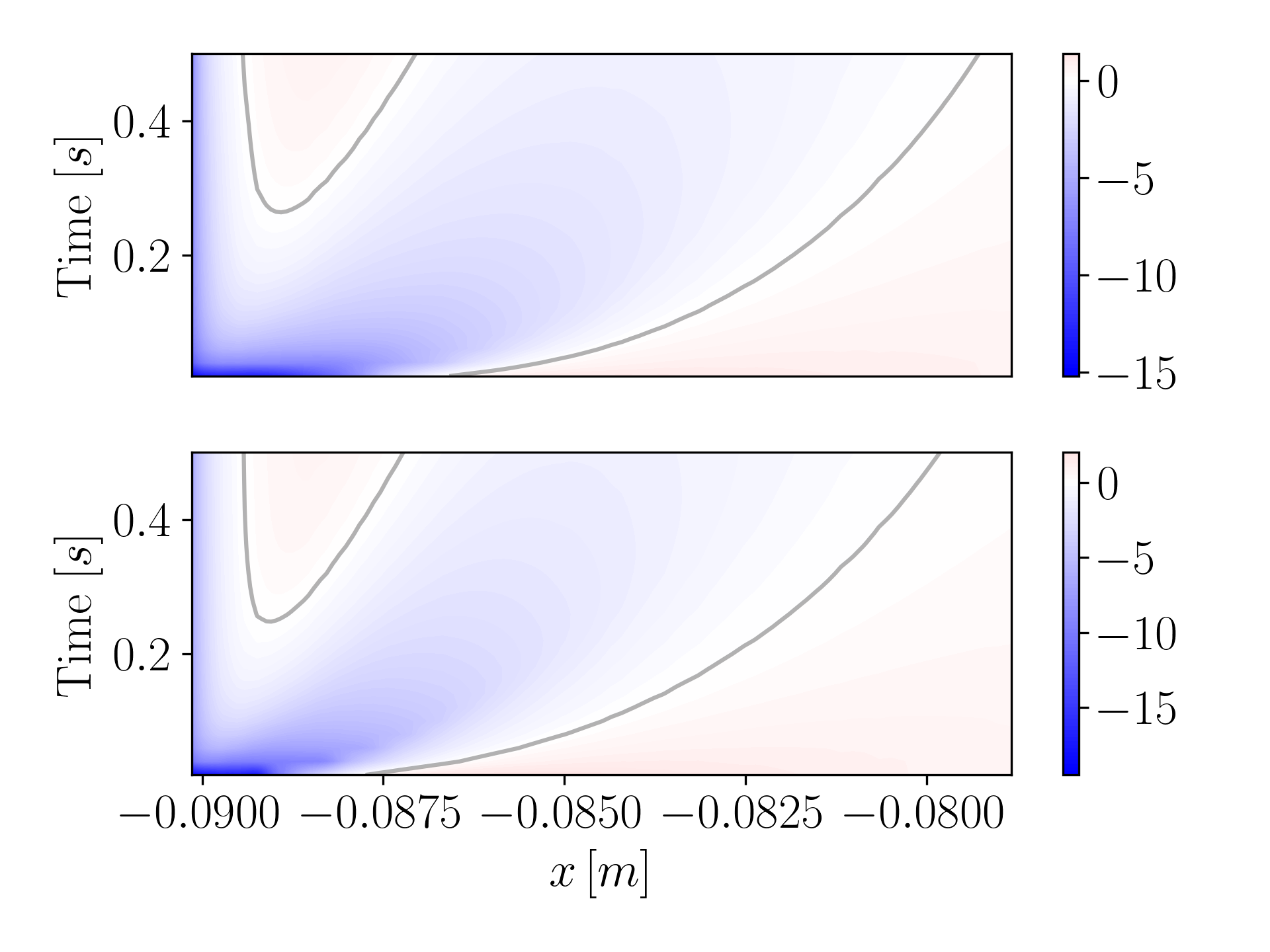}
\node at (0.5,1.03) {\small Case 4};
\node at (0.65,0.9) {\tiny New $B^\prime$};
\node at (0.65,0.48) {\tiny Classical $B^\prime$};
\end{tikzonimage}
\end{minipage}
\caption{Contour plot of the gas velocity (in  meters per second) normal to the surface. The black contour lines emphasize the $0$-contour, i.e., the transition from negative gas velocities (aspiration) to positive gas velocities (blowing).
The top panels are given by the new $B^\prime$ formulation, while the bottom panels by the classical $B^\prime$ formulation.}
\label{fig:contours}
\end{figure}

Figures \ref{fig:recession_uncoupled_t020} and \ref{fig:recession_uncoupled_t050} show the degree of surface recession at times $t = 0.20\,s$ and $t = 0.50\,s$, respectively, for the four different cases considered in table \ref{tab:cases}. 
The top half of all panels ($y \geq 0$) shows the wedge geometry as predicted by the new $B^\prime$ formulation, while the bottom half shows the geometry as given by the classical $B^\prime$ formulation. 
The geometry is colorcoded by the local instantaneous recession velocity \revision{(in meters per second)} normal to the surface. 
From the figures, we see that in the high pressure cases (cases 1 and 4), the new $B^\prime$ formulation predicts a higher recession velocity and a larger shape deformation. 
By contrast, in the low pressure cases (cases 2 and 3) the two formulations give (almost) identical predictions. 
These observations can be explained by looking at the time history of $\Bg$ at the leading edge of the wedge in figure \ref{fig:bg_uncoupled}. 
Here, we see that for cases 2 and 3 (panels \textit{(b)} and \textit{(c)}), $\Bg \geq 0$ for (almost) all times, meaning that porous material gases are blown into the boundary layer.
In this case, the two formulations are mathematically identical and it should therefore be expected that they predict the same surface recession velocities. 
On the other hand, we see that for cases 1 and 4 (panels \textit{(a)} and \textit{(d)}), $\Bg$ in the new formulation (solid lines) remains negative for all times, while $\Bg = 0$ in the classical formulation (dashed lines). 
By equation \eqref{eq:bc_3}, a negative $\Bg$ leads to a larger $\Bc$, which, in turn, gives higher recession velocities. 
\revisiontwo{Before moving forward, it is important to remark that in both $\Bg$ formulations, boundary layer gases are allowed to flow into the porous material (this can be seen clearly in figures \ref{fig:gas_vel_uncoupled} and \ref{fig:contours}). 
However, in the classical $\Bg$ formulation the effect of inflowing gases on the surface chemistry is neglected and $\Bg$ is not allowed to attain negative values.}

\revision{We now further investigate cases 1 and 4. 
While both cases exhibit sustained negative $\Bg$ values, we seek an explanation for the observation that, in case 4, there is a much more pronounced difference between the new and the classical $B^\prime$ formulations. 
This difference is evident from figure \ref{fig:recession_uncoupled_t050}d, where we see that the top surface (given by the new $B^\prime$ formulation) has receded almost twice as much as the bottom surface (given by the classical $B^\prime$ formulation).
Ultimately, as discussed throughout the manuscript, the reason behind the discrepancy between the two formulations is driven by 
\begin{equation}
    \Delta \Bg = B^\prime_{g,\text{classical}} - B^\prime_{g,\text{new}},
\end{equation}
which is significantly larger in case 4 (figure \ref{fig:bg_uncoupled}d) than in case 1 (figure \ref{fig:bg_uncoupled}a).} 

\begin{figure}
\centering
\begin{minipage}{0.48\textwidth}
\begin{tikzonimage}[trim= 15 100 5 90,clip,width=0.95\textwidth]{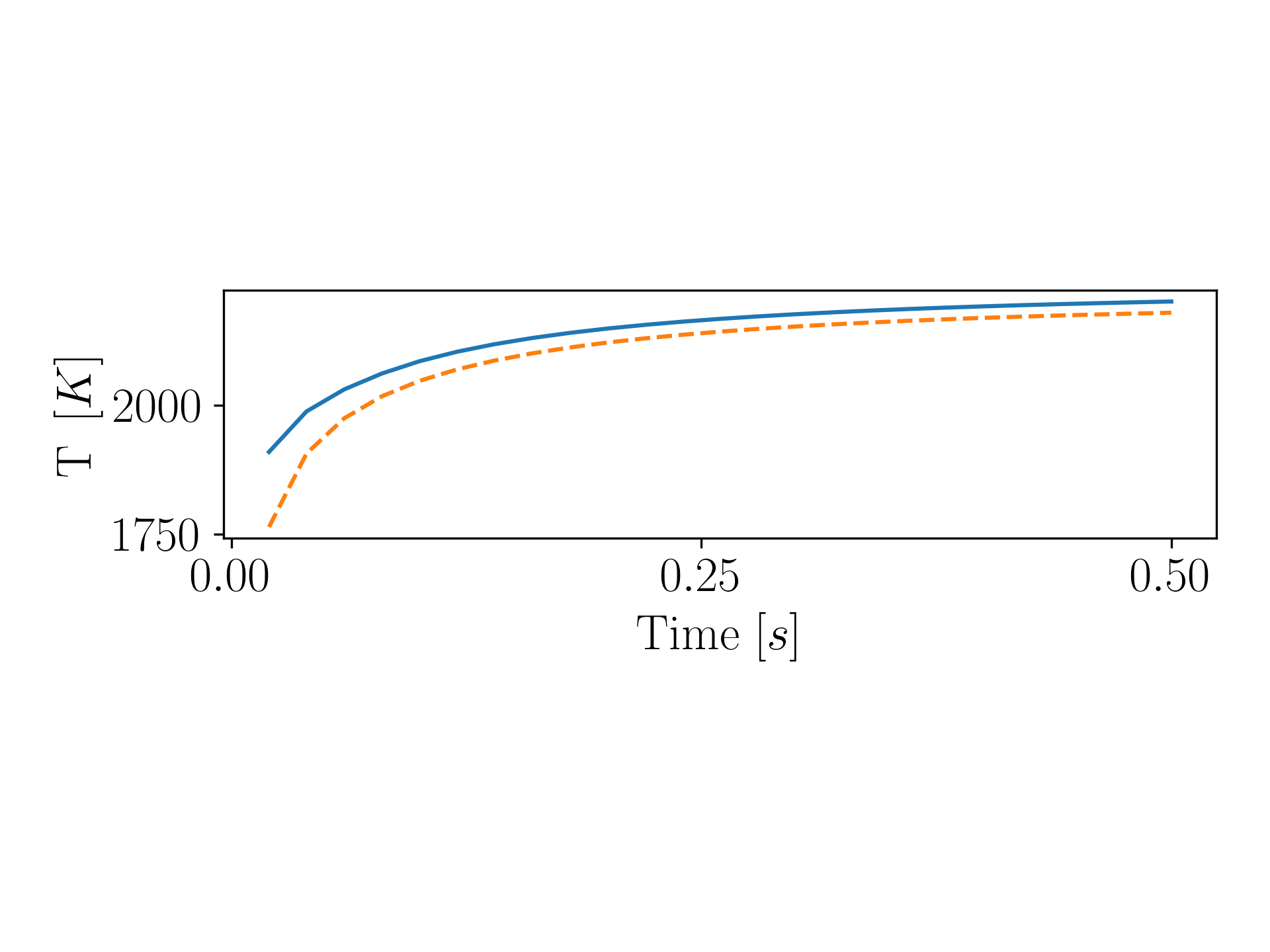}
\node at (0.9,0.45) {\small $\textit{(a)}$};
\end{tikzonimage}
\end{minipage}
\begin{minipage}{0.48\textwidth}
\begin{tikzonimage}[trim= 15 100 5 90,clip,width=0.95\textwidth]{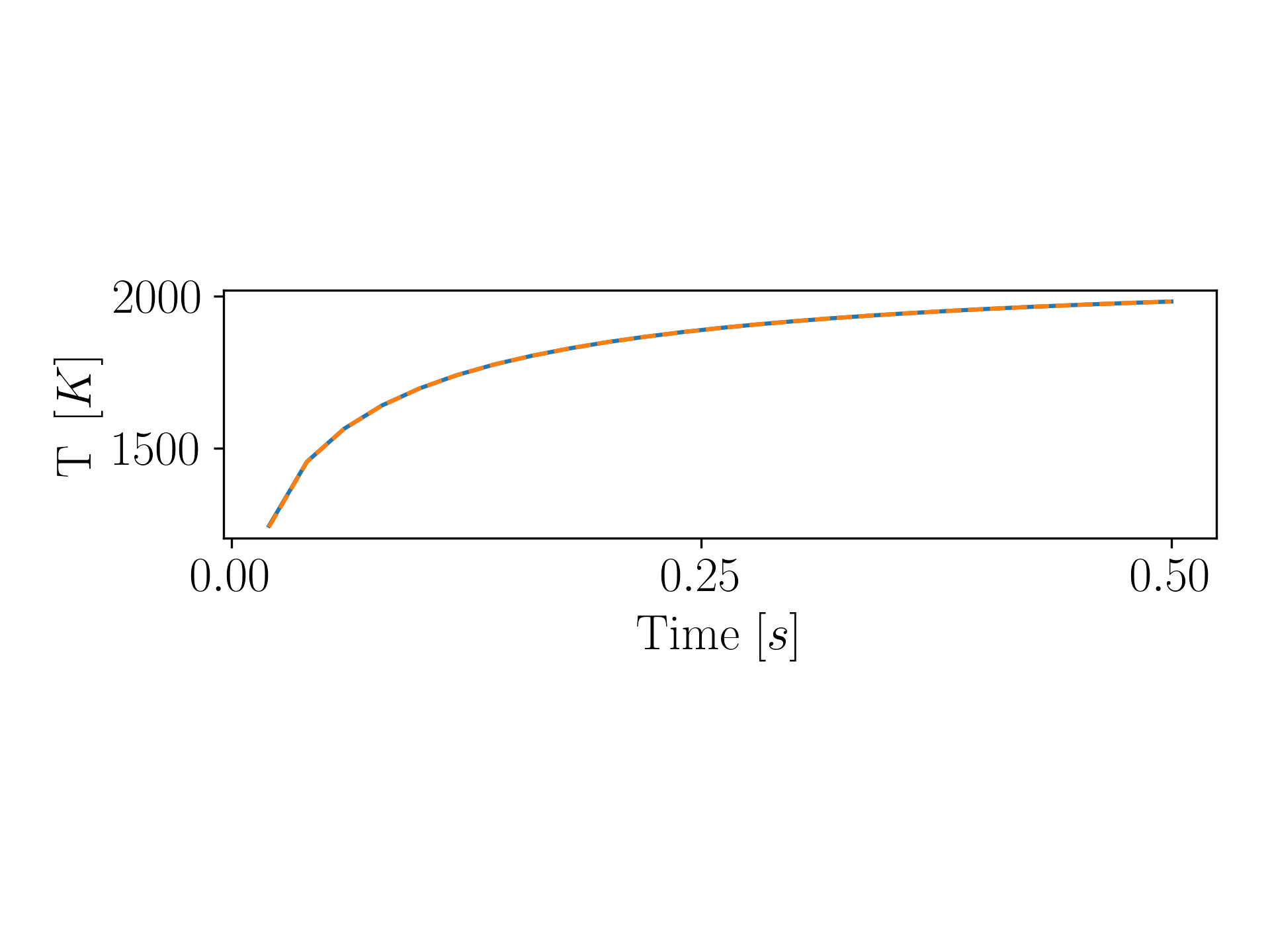}
\node at (0.9,0.45) {\small $\textit{(b)}$};
\end{tikzonimage}
\end{minipage}
\begin{minipage}{0.48\textwidth}
\begin{tikzonimage}[trim= 15 100 5 90,clip,width=0.95\textwidth]{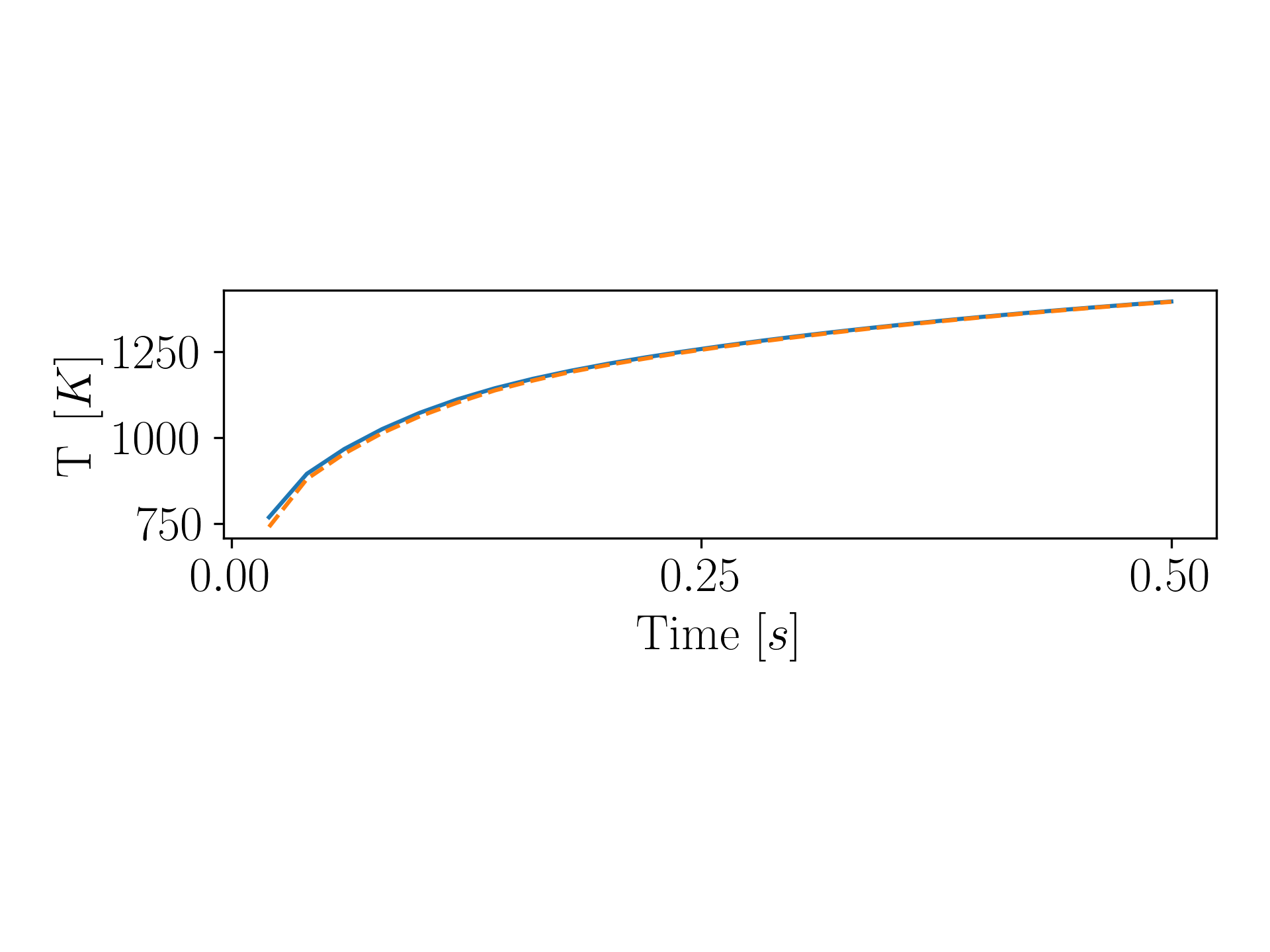}
\node at (0.9,0.45) {\small $\textit{(c)}$};
\end{tikzonimage}
\end{minipage}
\begin{minipage}{0.48\textwidth}
\begin{tikzonimage}[trim= 15 100 5 90,clip,width=0.95\textwidth]{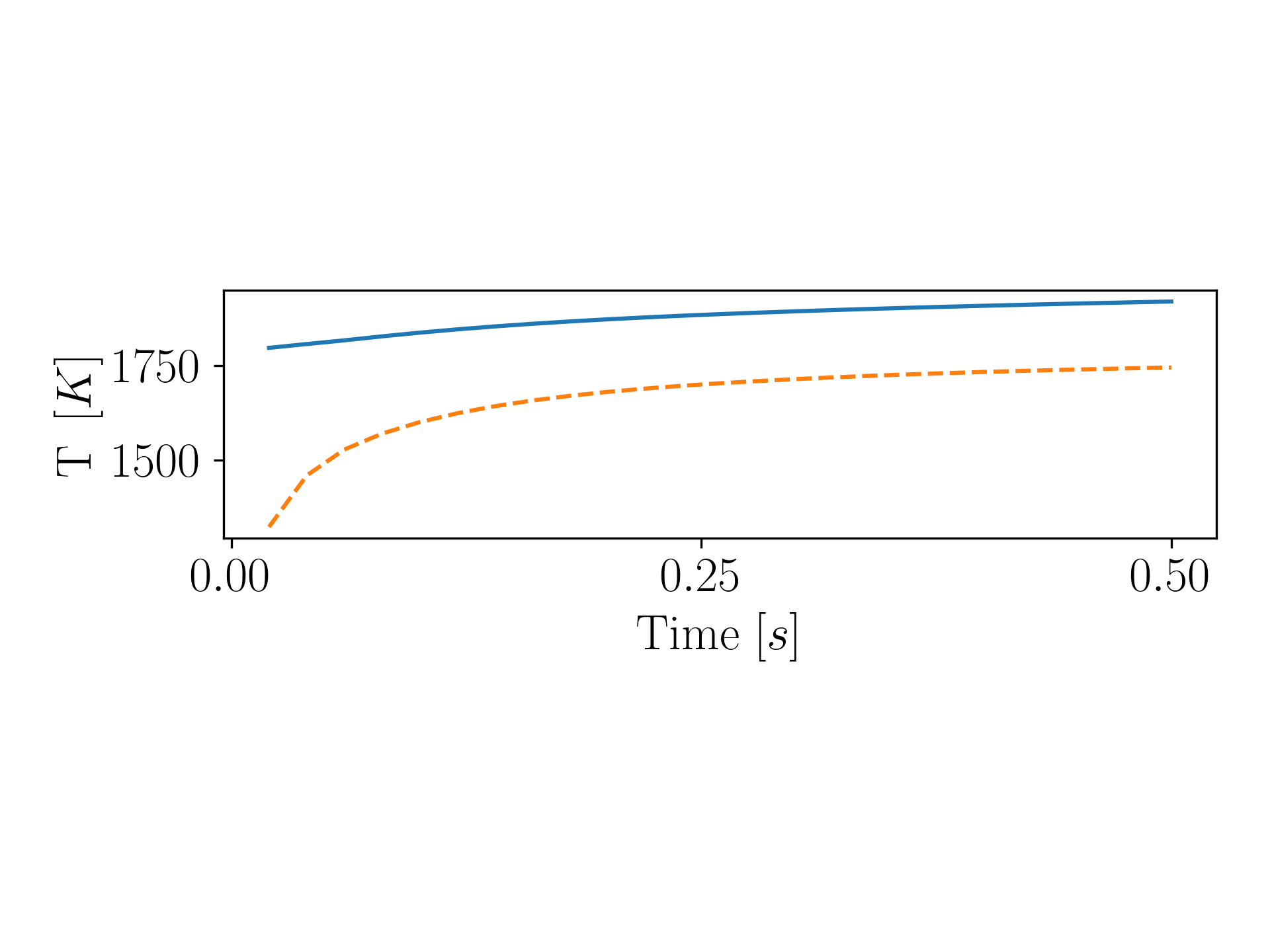}
\node at (0.9,0.45) {\small $\textit{(d)}$};
\end{tikzonimage}
\end{minipage}
\caption{Analog of figure \ref{fig:bg_uncoupled}, for the time-history of the surface temperature at the leading edge of the wedge.}
\label{fig:temp_uncoupled}
\end{figure}

\revision{In order to understand the difference between $\Delta \Bg$ in cases 1 and 4, we first recall that $\Bg$ can be understood as a normalized mass flux and, as such, it scales linearly with the local gas density and the local gas velocity. 
Interestingly, we see from figures \ref{fig:gas_vel_uncoupled}a and \ref{fig:gas_vel_uncoupled}d that the gas velocity at the stagnation point is approximately equal for both cases 1 and 4. 
(This is likely due to the fact that both cases are exposed to the same pressure boundary condition (see table \ref{tab:cases}).)
It follows that the difference in~$\Delta \Bg$ must be due to a proportional difference in the gas density, with a higher gas density in case 4 (thus, higher mass flux and larger $\Delta \Bg$) and a lower gas density in case 1. 
The reason why case 4 exhibits a higher gas density can be easily understood by recalling that case 4 is exposed to a normalized heat flux that is four times lower than that imposed in case 1 (see, once again, table \ref{tab:cases}).
Consequently, the temperature at the wedge leading edge in case 4 (figure \ref{fig:temp_uncoupled}d) is lower than its counterpart in case 1 (figure \ref{fig:temp_uncoupled}d), thereby leading to higher and lower densities, respectively.}

In light of this discussion, we conclude that aspiration ($\Bg < 0$) has a larger effect on the recession velocity at lower temperature and higher pressures. 
From an intuitive standpoint, the high pressure is necessary to cause aspiration (i.e., $\Bg < 0$), and this is required to observe any sort of difference between the two formulations. 
Clearly, the higher the pressure the higher the difference. 
However, as discussed, we also observe that the surface temperature has a non-negligible effect on the surface recession, with higher temperatures leading to higher recession velocities (case 1), but lower temperatures causing a larger spread $\Delta \Bg$ between the two formulations. 

\revision{In closing the results section, it is also interesting to study the inflow/outflow of gases into and out of the porous material as a function of time.
To do so, we focus on cases 3 and 4, and we plot contours of the gas velocity normal to the surface as a function of time and streamwise location along the wedge surface (figure \ref{fig:contours}).
In both cases, we do not observe noteworthy qualitative differences between the flow of gases computed using the new $B^\prime$ formulation (top panels) and the classical $B^\prime$ formulation (bottom panels). 
This suggests that accounting for the inflow of gases into the porous material has an effect primarily in the surface recession rate and in the surface thermodynamics (as discussed in the preceding paragraphs). 
Despite this, figure \ref{fig:contours} is still interesting, and it can be used to better understand the physics at hand. 
Interestingly, in case 3 we observe a ``flow reversal" whereby gases that are initially flowing into the material at early times and near the wedge leading edge, are eventually expelled along the whole surface at later times. 
(This is likely to be attributed to a rise in pressure inside the material due to pyrolysis, as discussed in \cite{lachaud2015}.)
Except for early times, $\Bg \geq 0$ along the whole surface, so the new $B^\prime$ formulation is mostly in agreement with the classical $B^\prime$ formulation, and the integrated difference in terms of surface recession is qualitatively negligible (see figures \ref{fig:recession_uncoupled_t020}c and \ref{fig:recession_uncoupled_t050}c). 
Case 4, on the other hand, exhibits much larger space-time regions of gas inflow, so it is to be expected that accounting for the effect of aspiration in the $B^\prime$ formulation will lead to significant differences in the predicted surface recession (see figures \ref{fig:recession_uncoupled_t020}d and \ref{fig:recession_uncoupled_t050}d). 
Interestingly, case 4 does not exhibit the same flow reversal as case 3, except for a narrow region on the wedge shoulder (approximately between $x = -0.089$ and $x = 0.087$ and after time $t \approx 0.25$).
}

\section{Conclusion}

We derive the $B^\prime$ formulation for ablating-surface boundary conditions from first principles,  starting from a jump condition that we obtained following the approach of \cite{keller1954}. 
This allows to clearly identify all the underlying assumptions of the $B^\prime$ formulation, especially when applied at a reacting interface between a boundary layer and a porous material.
We then extend the $B^\prime$ formalism to account for the advective transport of boundary layer gases into the porous material. 
Although this is a common occurrence in hypersonics applications and in thermal protection systems, the classical $B^\prime$ formulation neglects its effect on the dynamics of the material.
We demonstrate, both theoretically and via examples, that accounting for the advective transport of gases into the porous material can have a significant effect on the recession velocity of ablating interfaces. 


\section*{Acknowledgments}

This material is based upon work supported by the National Science Foundation under Grant No. 2139536, issued to the University of Illinois at Urbana-Champaign by the Texas Advanced Computing Center under subaward UTAUS-SUB00000545 with Dr. Daniel Stanzione as the PI.  
The computations were performed on TACC’s Frontera under LRAC grant CTS20006.

\appendix

\section{Generating the $B^\prime$ Tables}
\label{app:tables}

Here, we describe how the $B^\prime$ tables (for the new framework) can be generated using Mutation++ \citep{scoggins2020}.
We seek a table whose independent variables are the wall pressure $p$, the wall temperature~$\langle T\rangle$ and the normalized blowing rate $\Bg$ on the porous material's side of the interface. 
Given $p$, $\langle T\rangle $ and $\Bg$ as inputs, the tables will output (after interpolation, if necessary), the normalized recession rate $\Bc$ and the enthalpy $h_w$ of the equilibrium mixture. 

When generating the tables, some care is required. 
In particular, Mutation++ generates tables as a function of $p$, $\langle T\rangle$ and the normalized mass flux of species that are advected \emph{towards} the interface.
Depending on the specific case (see subsections below), this normalized mass flux is either $\Bg$ or $\Bf$. 
As mentioned, however, during computation we would like to perform table look-ups based on $p$, $\langle T\rangle$ and $\Bg$, since $\Bg$ is a quantity that is always readily computed by the material response solver (recall the definition of $\Bg$ from equation \eqref{eq:bprime}). 
In order to be able to perform look-ups based on $p$, $\langle T\rangle$ and $\Bg$, the tables generated by Mutation++ require some post-processing.

\subsection{Table for $\Bg \geq 0$}
From section \ref{sec:extension_bprime}, this case corresponds to $\Bf \geq \Bc$.
This table can be generated using Mutation++ directly, without any further post processing, since the normalized mass flux of species that are advected towards the interface is precisely $\Bg$ .
The composition of the reactants used for the equilibrium calculations is speciefied in section \ref{subsec:case1}.
Henceforth, we refer to this table as Table I. 

\subsection{Table for $\Bf < 0$}
From section \ref{sec:extension_bprime} this is one of the two cases corresponding to $\Bg < 0$. (The other case is $0\leq \Bf \leq \Bc$, discussed shortly.)
This table can also be generated using Mutation++, with the reactants composition specified in section \ref{subsec:case3}.
However, the normalized mass flux used by Mutation++ corresponds to $\Bf$ (and not $\Bg$, as desired). 
Fortunately, by mass conservation, we know that $\Bg = \Bf - \Bc$. 
The table generated by Mutation++ can then be easily rearranged such that the look-up can be performed based on $\Bg$.
We henceforth refer to this table as Table~II. 

\subsection{Table for $0\leq\Bf \leq \Bc$}
This is the other case corresponding to $\Bg < 0$. 
However, we recall from section \ref{subsec:case2}, that in this specific case $\Bc$ and $h_w$ (i.e., the outputs of the $B^\prime$ tables) are independent of $\Bf$ or $\Bg$.
Then, for a given $p$ and $\langle T\rangle$, the outputs $\Bc$ and $h$ can be calculated from Table I with $\Bg = 0$. 
We henceforth refer to this table as Table~III.
Finally, a unified $B^\prime$ table can be obtained by ``stacking" together tables II, III and~I (in increasing $\Bg$ order, from negative to positive). 

\section{Mass- and Heat-Transfer Boundary Layer Analogy}
\label{app:St}

While this topic is addressed in \cite{eckert1969} and \cite{incropera2007}, and touched upon in \cite{cooper2022} and in Appendix A in \cite{meurisse2018}, we repropose the derivation of the mass- and heat-transfer boundary layer analogy. 
This will clarify the definition of mass- and heat-transfer Stanton numbers, as well as the interpretation of the mass- and heat-transfer potential models used in the $B^\prime$ mass and energy balances.  

Following \cite{eckert1969}, we begin with the steady, zero-pressure-gradient boundary layer equations
\begin{align}
    \frac{\partial \rho u}{\partial x} + \frac{\partial\rho v}{\partial y} &= 0 \label{eq:continuity_bl}\\
    \rho u \frac{\partial u}{\partial x} + \rho v \frac{\partial u}{\partial y} &= \frac{\partial}{\partial y}\left(\mu \frac{\partial u}{\partial y}\right)\label{eq:momentum_bl} \\
    \rho u \frac{\partial H}{\partial x} + \rho v \frac{\partial H}{\partial y} &= -\frac{\partial \varepsilon}{\partial y} +\frac{\partial}{\partial y}\left(\mu u\frac{\partial u}{\partial y}\right)\label{eq:enthaply_transport} \\ 
    \rho u \frac{\partial w_i}{\partial x} + \rho v \frac{\partial w_i}{\partial y} &= -\frac{\partial j_i}{\partial y}.\label{eq:species_transport}
\end{align}
Here, $w_i$ are the mass fractions in a mixture with $N_s$ species, $H = h + (1/2)(u^2 + v^2)$ is the total enthalpy and the fluxes $\varepsilon$ and $j_i$ are defined as 
\begin{equation}
\label{eq:fluxes}
    \varepsilon = -\kappa \frac{\partial T}{\partial y} + \sum_{i=1}^{N_s} h_i j_i,\quad j_i = -\rho D_i \frac{\partial w_i}{\partial y}. 
\end{equation}
The definition of $j_i$ is known as Fick's law, with diffusion coefficient $D_i$ associated with species $i$. 
The definition of $\varepsilon$, on the other hand, is the sum of Fourier's law for heat conduction, and the transport of enthalpy due to diffusion (see, e.g., \cite{ramshaw2002}). 

Using the definition of the fluxes in \eqref{eq:fluxes}, equations \eqref{eq:species_transport} and \eqref{eq:enthaply_transport} can be cast in conservative form using \eqref{eq:continuity_bl} and \eqref{eq:momentum_bl},
\begin{align}
    \frac{\partial }{\partial x}\left(\rho u h\right) + \frac{\partial }{\partial x}\left(\rho v h + \varepsilon\right) &= 0, \label{eq:enthalpy_transport_conservative}\\
    \frac{\partial}{\partial x}\left( \rho u w_i \right) + \frac{\partial}{\partial y}\left(\rho v w_i - \rho D_i \frac{\partial w_i}{\partial y}\right) &= 0 \label{eq:species_transport_conservative}
\end{align}
In obtaining \eqref{eq:enthalpy_transport_conservative} we have used the definition of $H$, neglected the term $\rho u \partial v/\partial x + \rho v \partial v/\partial y$ (consistently with the scaling arguments that led to the velocity boundary layer equation \eqref{eq:momentum_bl}), and neglected the viscous dissipation term $\mu \left(\partial u/\partial y\right)^2$ (see page 366 in \cite{incropera2007}). 

For boundary layer analogy between the thermal boundary layer \eqref{eq:enthalpy_transport_conservative} and the species boundary layer \eqref{eq:species_transport_conservative}, we require 
\begin{equation}
\label{eq:required_equality}
    \varepsilon = - \rho D_i \frac{\partial h}{\partial y}.
\end{equation}
As a first step, we observe that the enthalpy of the mixture can be expressed as 
\begin{equation}
    h(T,w) = \sum_{i=1}^{N_s} h_i(T)w_i,
\end{equation}
so that, using the chain rule and defining $c_p = \partial h/\partial T$, we have
\begin{equation}
    dT = \frac{1}{c_p} dh - \frac{1}{c_p}\sum_{i=1}^{N_s} h_i(T) dw_i.
\end{equation}
Using the definition of $\varepsilon$ in \eqref{eq:fluxes} and the equation above, we can write 
\begin{equation}
\label{eq:eps_intermediate}
    \varepsilon = -\frac{\kappa}{c_p} \frac{\partial h}{\partial y} + \frac{\kappa}{c_p}\sum_{i=1}^{N_s}h_i \frac{\partial w_i}{\partial y} - \rho \sum_{i=1}^{N_s} h_i D_i \frac{\partial w_i}{\partial y}.
\end{equation}
Defining the Prandlt and Schmidt numbers 
\begin{equation}
    Pr = \frac{\mu c_p}{\kappa},\quad Sc_i = \frac{\mu}{\rho D_i},
\end{equation}
we can write \eqref{eq:eps_intermediate} as 
\begin{equation}
    \varepsilon = -\frac{\mu}{Pr} \frac{\partial h}{\partial y} + \frac{\mu}{Pr} \sum_{i=1}^{N_s}\left(1 - \frac{Pr}{Sc_i}\right)h_i\frac{\partial w_i}{\partial y}.
\end{equation}
From this equation, it is immediate that \eqref{eq:required_equality} is satisfied so long as $Pr = Sc_i$ (i.e., if the species Lewis number $Le_i = Pr/Sc_i$ is equal to $1$).
Thus, given the set of assumptions made throughout this derivation, mass- and heat-transfer boundary layer analogy is achieved for species Lewis numbers $Le_i = 1$. 
We note in passing that to achieve analogy with the velocity boundary layer in \eqref{eq:momentum_bl}, one would also require $Pr = 1$. 
Before moving forward, we wish to point out that the derivation of the boundary layer analogoy presented herein is slightly different than the one in \cite{eckert1969}, where the author worked directly with total enthalpy. 
This led to a different set of assumptions and to the additional requirement of $Pr$ for thermal/species boundary layer analogy. 

Using the derivation above, we can now straightforwardly define the mass-transfer and heat-transfer Stanton numbers.
Assuming equal diffusion coefficients $D = D_i$ for all $i$, the mass-transfer Stanton number $St_M$ is defined as 
\begin{equation}
\label{eq:StM_def}
    j_i = - \rho D \frac{\partial w_i}{\partial y} \coloneqq \rho_e u_e St_M\left(w_{i,s} - w_{i,e}\right),
\end{equation}
where the subscript ``e" denotes an edge quantity and the subscript ``s" denotes a surface quantity. 
The heat-transfer Stanton number $St_H$ is defined similarly,
\begin{equation}
\label{eq:StH_def}
    \varepsilon = \rho_e u_e St_H \left(h_s - h_e\right).
\end{equation}
By the aforementioned boundary layer analogy, it follows immediately that 
\begin{equation}
    St \coloneqq St_M = St_H. 
\end{equation}

As a final note, it is interesting to express the contribution of $\kappa \partial T/\partial y$ to $\varepsilon$ in terms of the Stanton number. 
Starting from the definition of $\varepsilon$ in \eqref{eq:fluxes}, using \eqref{eq:StM_def} and \eqref{eq:StH_def} alongside the boundary layer analogy and equal diffusion coefficients, we have
\begin{equation}
    \varepsilon = \rho_e u_e St \left(h_s - h_e\right) = -\kappa \frac{\partial T}{\partial y} + \rho_e u_e St \underbrace{\sum_{i=1}^{N_s}h_i\left(w_{i,s} - w_{i,e}\right)}_{\coloneqq \left(h_s - h_{s,e}\right)},
\end{equation}
which implies 
\begin{equation}
    -\kappa \frac{\partial T}{\partial y} = \rho_e u_e St \left(h_{s,e} - h_e\right),
\end{equation}
where $h_{s,e}$ is the enthalpy at the surface with edge composition.

\bibliographystyle{elsarticle-harv} 
\bibliography{references}

\end{document}